
\hsize 6.0 true in
\vsize 9.0 true in

\voffset =  -.2 true in
\font\tentworm=cmr10 scaled \magstep2
\font\tentwobf=cmbx10 scaled \magstep2

\font\tenonerm=cmr10 scaled \magstep1
\font\tenonebf=cmbx10 scaled \magstep1

\font\eightrm=cmr8
\font\eightit=cmti8
\font\eightbf=cmbx8
\font\eightsl=cmsl8
\font\sevensy=cmsy7
\font\sevenm=cmmi7

\font\twelverm=cmr12
\font\twelvebf=cmbx12
\def\subsection #1\par{\noindent {\bf #1} \noindent \rm}

\def\mid {\let\rm=\tenonerm \let\bf=\tenonebf \rm \bf}

\def\para{\par \vskip 12 pt}

\def\head{\let\rm=\tentworm \let\bf=\tentwobf \rm \bf}

\def\heading #1 #2\par{\centerline {\head #1} \smallskip
 \centerline {\head #2} \vskip .15 pt \rm}

\def\eight{\let\rm=\eightrm \let\it=\eightit \let\bf=\eightbf
\let\sl=\eightsl \let\sy=\sevensy \let\m=\sevenm \rm}

\def\foots{\noindent \eight \baselineskip=10 true pt \noindent \rm}
\def\sexion{\let\rm=\twelverm \let\bf=\twelvebf \rm \bf}

\def\section #1 #2\par{\vskip 20 pt \noindent {\mid #1} \enspace {\mid #2}
  \para \noindent \rm}

\def\ssection #1 #2\par{\noindent {\mid #1} \enspace {\mid #2}
  \para \noindent \rm}

\def\abstract#1\par{\para \foots {\bf Abstract: \enspace}#1 \para}

\def\author#1\par{\centerline {#1} \vskip 0.1 true in \rm}

\def\abstract#1\par{\noindent {\bf Abstract: }#1 \vskip 0.5 true in \rm}

\def\midsection #1\par{\noindent {\sexion #1} \noindent \rm}

\def\sqr#1#2{{\vcenter{\vbox{\hrule height#2pt
 \hbox {\vrule width#2pt height#1pt \kern#1pt
  \vrule width#2pt}
  \hrule height#2pt}}}}

\baselineskip=18pt


\def\n{\noindent}
\def\s{\smallskip}
\def\m{\medskip}
\def\b{\bigskip}


\def\mn{{\mu\nu}}
\def\m{\mu}
\def\a{\alpha}
\def\b{\beta}
\def\l{\lambda}
\def\d{\delta}
\def\o{\omega}
\def\g{\gamma}
\def\t{\theta}
\def\p{\partial}
\def\L{\Lambda}
\def\D{\Delta}
\def\O{\Omega}
\def\G{\Gamma}
\def\ab{{\alpha \beta}}

\def\mnl{{\mu \nu \lambda}}
\def\abg{{\alpha \beta \gamma}}
\def\agd{{\alpha \gamma \delta}}
\def\boxit#1{\vbox {\hrule \hbox {\vrule \kern3pt
\vbox{\kern3pt#1\kern3pt}\kern3pt\vrule}\hrule}}
\def\boxit#1{\vbox {\hrule \hbox {\vrule \kern3pt
\vbox{\kern3pt#1\kern3pt}\kern3pt\vrule}\hrule}}

\def \dal {\boxit {\phantom ,}}

\line{\hfill \foots{\bf IUCAA Preprint}}
\line{\hfill \foots{ Jan. 1992}}
\vskip .45 true in

\heading{QUANTUM EFFECTS}

\heading{NEAR A POINT MASS IN }

\heading{2+1 DIMENSIONAL GRAVITY}

\vfill
\centerline{by}
\smallskip
\centerline{Tarun Souradeep$^*$ and Varun Sahni$^{\dag}$}
\medskip
\centerline{Inter-University Centre for Astronomy and Astrophysics}
\centerline{Post Bag 4, Ganeshkhind, Pune 411 007}
\centerline{INDIA}

\vfill
\settabs 10\columns
\+&&&email~:&$^*$~~~~tarun@iucaa.ernet.in \cr
\+&&&&$\dag$~~~~varun@iucaa.ernet.in\cr
\medskip

$$ To ~~appear~~in ~{\bf \rm Physical ~Review ~D} $$
\eject



\section{Abstract}

We investigate the behaviour of classical and quantum fields in the conical
space-time
associated with a point mass in 2+1 dimensions. We show that the presence of
conical boundary conditions alters the electrostatic field of a point charge
leading
to the presence of a finite self-force on the charge from the direction of the
point mass exactly as if the point mass itself were charged. The conical
space-time
geometry also affects the zero point fluctuations of a quantum scalar field
leading to
 the existence of a vacuum polarisation --- $\langle T_\mn \rangle$, in the 2+1
dimensional analog of the Schwarzschild metric. The resulting linearised
semi-classical
Einstein equations --- $G_\mn = 8 \pi G \langle T_\mn \rangle$, possess a well
defined Newtonian limit, in marked contrast to the classical case for which no
Newtonian limit is known to exist. An elegant reformulation
of our results in terms of the method of images is also presented.

\n Our analysis also covers the non-static de Sitter--Schwarzschild metric in
2+1
dimensions, in which in addition to the vacuum polarisation, a non-zero vacuum
flux of
energy -- $\langle T_{rt} \rangle $ is also found to exist.As part of this
analysis,
we evaluate the scalar field propagator in an $n$-dimensional de Sitter space,
as a
result some novel features of quantum field theory in odd dimensions are seen
to emerge.

$$ {\bf PACS ~~Nos.} - 4.2 ,~4.6 , ~11.10-z , ~03.70+k ,~98.80-k.$$

\vfill\eject


\section{1. Introduction}

A well established feature of Einstein gravity is that in space-time of
dimensionality $d < 4$, it is devoid of any intrinsic dynamics. In 3 dimensions
this result arises from the observation that both the Ricci and the Riemann
tensors have an equal number of components (=6). Consequently, the Riemann
tensor can be expressed in terms of a
combination of Ricci tensors~:

$$R^\mn_\ab = \epsilon^\mnl \epsilon_\abg ( R_\lambda^\gamma -
 {1\over 2} \d^\gamma_\lambda R).\eqno(1.1)$$

\n Clearly if the Ricci tensor vanishes then so does the Riemann tensor,
with the result that gravity does not propagate outside of matter sources.
Since the Weyl tensor incorporates the internal degrees of freedom of the
gravitational field it
follows from (1.1) that $C_{iklm} \equiv 0$ in a 3 dimensional space-time.
In space-time of dimensionality $d > 3$, the vanishing of the Weyl tensor
(also called`the conformal tensor') is indicative of the fact that the
space-time
 under consideration is conformally flat. This is not so in $d = 3$,
the issue of conformal
flatness in this case being decided not by the Weyl tensor but by the
symmetric, conserved and traceless Cotton-York tensor - (sometimes also
known as the three dimensional Weyl tensor)

$$C^\ab = \epsilon^\agd ( R^\b_\g - {1\over 4} \d^\b_\g R );_\d ,\eqno(1.2)$$

\n so that any three dimensional space-time is conformally flat if and only if
the
Cotton-York tensor vanishes$^{1}$. The Cotton-York tensor also features
prominently
in topologically massive gravity which has been the focus of considerable
attention in recent years following the discovery by Deser that bosons and
fermions
can acquire exotic spin and statistics within the framework of this
theory$^{2,3}$.

\n Topologically massive gravity is described by an action which is the sum of
 the standard Einstein action and a Chern-Simons term$^3$

$$I = I_E + {1 \over \m} I_{cs} \eqno(1.3a)$$

\n where

$$I_E = \int d^3x \sqrt{g} R,\eqno(1.3b)$$

\n and $I_{cs}$ is the Chern-Simons action

$$I_{cs} = {1\over 2\m} \int d^2x \sqrt{g} \epsilon^{\mu \alpha \nu }
\bigg[ \omega^a_\m \partial_\alpha \omega_{\n a} + {1\over 3} \omega^a_\m
\o^b_\a
\o^c_\nu \epsilon_{abc}\bigg]\eqno(1.3c)$$

\n where $\o^a_b$ is the spin connection and $\m$ is a constant having
dimension of mass.

\n Variation of $I$ with respect to the metric results in the Einstein-Cotton
equations$^{2,3}$

$$R^\mn + {1\over \m} C^\mn = 0\eqno(1.4)$$

\n $C^\mn$ being the Cotton-York tensor.

\n The new equations of motion (1.4), do not constrain the curvature to vanish
in the
absence of sources, so that gravity has a nontrivial dynamics and can
propagate.

\n It is interesting to note that the
external metric of a static point source is identical in both topologically
massive
gravity and Einstein gravity at large distances from the source $(\mu r > 1)$,
and
is given by$^{2-6}$

$$ds^2 = dt^2 - dr^2 - r^2 d\varphi^2 \eqno(1.5a)$$

\n where

$$0 \leq \varphi \leq {2\pi \over p},~~~~p = (1 - 4 G_2 M)^{-1},~~(p \geq 1),$$

\n $M$ being the mass of the point source and $G_2$ -- the gravitational
constant
 in 2+1 dimensions.

\n We note that the $t =~ constant~$ 2-space of this metric is a cone,
and that the metric is flat everywhere except at the origin.

\n In terms of a new polar coordinate $\tilde \varphi = p \varphi$ the metric
takes the form

$$ds^2 = dt^2 - dr^2 - {r^2 \over p^2} d\tilde\varphi^2 \eqno(1.5b)$$

\n with $\tilde \varphi$ extending over the entire range $0 \leq \tilde \varphi
< 2 \pi$.

\n Metric (1.5) can be obtained from the well known exterior metric of a
straight
cosmic string by suppressing the dimension along its length$^7$.
Recently many
authors$^8$ have studied the scattering of point particles in the conical
space-time metric (1.5). We shall follow an alternate approach and study the
semi-classical one loop quantum gravitational effects that arise in such a
space-time due to its nontrivial topology. Such effects are also known to be
associated with cosmic strings and have been extensively studied by a number
of authors$^9$.

\n The outline of this paper is as follows~:

\n In $\S$2 we study the classical electrostatic field of a charged particle
in the space-time of a point mass (1.5). We show that the existence of boundary
conditions distorts the electrostatic field of the particle in a way that
causes
the particle to experience a repulsive self-force directed away from the point
mass.

\n In $\S$3 we examine quantum fluctuation of a massless scalar field in the
conical
space-time described by (1.5). We demonstrate the existence of a vacuum
polarisation characterised by a finite vacuum expectation value of the
energy-momentum
tensor --- $\langle T_\mn \rangle$. We also show that the vacuum energy density
---
$\langle T_{oo} \rangle$ is negative, for scalar fields coupling either
conformally or
minimally to gravity.

\n In $\S$4 we calculate the back reaction of one-loop quantum gravitational
effects
on the space-time geometry via the semi-classical Einstein equations $G_\mn =
8\pi G_2
\langle T_\mn \rangle$. We find that in the linearised approximation the
semiclassical
Einstein equations have a well defined Newtonian limit in marked contrast to
the
classical case where no such limit exists.

\n In $\S$5 we extend our study to the Schwarzschild-de Sitter metric. We find
in this case,
in addition to the vacuum polarisation
the presence of a vacuum energy flux $\langle T_{tr} \rangle$ directed
 radially away from the
point source.

\n In $\S$6 we extend our analysis to include twisted scalar fields and
evaluate
$\langle \phi^2\rangle^T$ and $\langle T_\mn \rangle^T$ for twisted fields in
the
three dimensional Schwarzschild metric. We find that $\langle \phi^2 \rangle^T$
and $\langle T_\mn \rangle^T$ are generally of opposite sign to $\langle \phi^2
\rangle$ and $\langle T_\mn \rangle$ for untwisted fields.

\n We end our paper with a discussion of our results in $\S$7.

\bigskip


\section{2. Electrostatics in 2+1 dimensions}

\n We begin our treatment of classical and quantum effects in conical
space-times
 with a study of the classical Poisson equation in the conical background
geometry (1.5).
Since fields are generally sensitive to the global properties of a space-time
one
would in general expect non-trivial modifications to arise to the standard
 electrostatic field of a point charge in (1.5).

\n A general solution to the Poisson equation

$$\D \varphi (\vec x) = -2 \pi q ~ \d^2 (\vec x - \vec x~')\eqno(2.1)$$

\n for a point charge located at $\vec x$ may be found by first constructing
a Green's function, satisfying

$$\D G (\vec x, \vec x~') = -2\pi ~\d^2 (\vec x - \vec x~').\eqno(2.2)$$

\n The self-force on a test charge in the space-time (1.5)
is then given by $\vec F = - \vec \bigtriangledown U (\vec x),$
where $U(\vec x)$ is the electrostatic energy,

$$U(\vec x) = {q^2 \over 2} G(\vec x, \vec x) .\eqno(2.3)$$

\n The symmetry of the problem makes it convenient to work in polar
coordinates, the
Poisson equation for the Green's function (2.2) then assumes the form

$$\bigg( {1\over r} {\partial \over \p r } r  {\partial \over \partial r} +
{1 \over r^2} {\partial^2 \over \partial \theta^2}\bigg) G_p (r, \theta; r',
\theta')
= - {2\pi \over r} \d(r - r') \d (\theta - \theta')$$

\n where $\d (r - r')$ and $\d (\theta - \theta')$ are one dimensional delta
functions.
Since $\d(\theta - \theta') = {p \over 2\pi} \sum\limits^\infty_{m = -\infty}
e^{ipm(\theta -\theta')}$, we shall use the polar coordinate
expansion of the Green's function

$$G_p (r, \theta; r', \theta') =  {p \over 2\pi} \sum\limits^\infty_{m =
-\infty}
e^{ipm(\theta -\theta')} g_m (r, r').\eqno(2.5)$$

\n $g_m (r, r')$ then satisfies the radial differential equation

$$\bigg( {\partial \over \partial r} r {\partial \over \partial r} - {p^2 m^2
\over r}
\bigg) g_m (r, r') = - 2\pi \d(r - r').\eqno(2.6)$$

\n Our one dimensional Green's function can be written as

$$\eqalign{g_m (r, r') & = -{1\over A} u_1 (r) u_2 (r') ~~~~~~~~~~r < r'\cr
&= -{1\over A} u_1 (r') u_2 (r) ~~~~~~~~~~r > r'\cr}\eqno(2.7)$$

\n where $u_{1,2}$ are solutions of the corresponding homogeneous equation~:
$u_1(r) \equiv r^{\vert pm\vert}$ and $u_2 (r) \equiv r^{- \vert pm \vert}.$

\n The constant $A$ may be determined from the Wronskian condition

$$u_1 (r) {d\over dr} u_2(r) - u_2(r) {d\over dr} u_1 (r) = {A\over
r},\eqno(2.8)$$

\n which gives $A = -2 pm$, so that finally

$$g_m (r, r') = {1\over 2 \vert pm \vert }  X^{\vert p m \vert}~~~~~~~(m \not=
0)
\eqno(2.9a)$$

\n where

$$\eqalign{ & X = {r \over r'} ~~~~~~~~~~~{\rm for}~ r' > r\cr
	     & X = {r' \over r} ~~~~~~~~~~~{\rm for}~ r'< r\cr}	\eqno(2.9b)$$
\n and

$$\eqalign{ & g_o (r,r') = - {\rm ln}~r' ~~~~~~0 \leq r < r'\cr
	     & g_o (r,r') = - {\rm ln}~r ~~~~~0 \leq r'< r.\cr }\eqno(2.9c)$$

\n The two-dimensional Green's function $G_p (x, x')$ now assumes the form

$$G_p (r, \theta; r' \theta' ) = {1\over 2\pi} \sum\limits^\infty_{m=1} {X^{pm}
\over
m}~{\rm cos}~pm (\theta - \theta') - {p \over 2\pi}~{\rm ln}~r' \eqno(2.10)$$

\n where $X = {r\over r'} < 1$ is assumed.

\n Performing the summation in (2.10) we finally get$^{12}$

$$G_p (r, \theta; r', \theta') = - {1 \over 4\pi} ln
\bigg[ r^{2p} + r'^{2p} - 2(rr')^p~{\rm cos}~p (\theta -
\theta')\bigg]\eqno(2.11)$$

\n which for $p = 1$ reduces to

$$G_1 (r, \theta ; r', \theta') = -{1\over 2\pi}~{\rm ln}~ \vert \vec x - \vec
x~' \vert
\eqno(2.12)$$

\n the familiar form for the Green's function in 2+1 dimensional Minkowski
space.
$G_p (\vec x, \vec x~')$ is formally divergent in the limiting case
$\vec x \rightarrow \vec x~'$, and must be regularised. Subtracting the flat
space
 contribution
$G_1 (\vec x, \vec x~')$ from $G_p (\vec x, \vec x~')$ and taking the limit
$\vec x \rightarrow \vec x~'$ we get

$$\eqalign{G^{\rm reg}_p (\vec x, \vec x) &= \lim_{\vec x \rightarrow \vec x~'}
 \bigg[ G_p (\vec x, \vec x~')
- G_1 (x, x') \bigg] \cr
&=-{1\over 4\pi}~{\rm ln}~[ p^2 r^{2(p-1)}],\cr}\eqno(2.13)$$
\n which is finite.

\n The electrostatic energy of a charge distribution is

$$U = {1\over 2} \int \int \rho (\vec x~') G_p^{\rm reg} (\vec x~', \vec x~'')
\rho (\vec x~'')
d^2 x' d^2 x''\eqno(2.14)$$

\n where $\rho (\vec x)$ is the charge density. For a point charge located at
$\vec x, \rho (\vec x) = q \d^2 (\vec x~' - \vec x)$, so that

$$\eqalign{U(\vec x) & = {q^2\over 2} \int \int \d^2 (\vec x~' - \vec x) \d^2
(\vec x~'' - \vec x)
G_p^{\rm reg} (\vec x~' , \vec x~'') d^2 x' d^2 x''\cr
&= {q^2 \over 2} G_p^{\rm reg} (\vec x, \vec x) = {- q^2 \over 8\pi}~{\rm ln}~[
p^2  r^{2(p-1)}
].\cr}\eqno(2.15)$$

\n The self-force on the test charge is then

$$\vec F = - \vec \bigtriangledown U = - \hat r {\partial U \over \partial r}
= \hat r {(p -1) q^2 \over 4\pi r}.\eqno(2.16)$$

\n We find that the self-force is repulsive and can be fairly large for
 $p \gg 1$, corresponding to large values of the deficit angle.$^{10}$

\n Interestingly, for $M \ll (4G_2)^{-1}$

$$\vec F \simeq {(4G_2 M q) q  \over 4\pi r}\hat r \equiv {Qq \over 4\pi r}
\hat r \eqno(2.17)$$

\n i.e., the conical boundary conditions present in (1.5) have effectively
induced a charge $Q$
on the point source, proportional to its mass and of the same sign as the
test charge $q$.

\n Our results can be elegantly rederived using the method of images$^{11}$,
 according to which
for integer $p ~(p \geq 1)$

$$\eqalign{G^{\rm reg}_p (r, \theta; r', \theta') &= \sum\limits^{p-1}_{k =1}
G_1 (r, \theta; r', \theta' +{2 \pi k \over p})\cr
&= -{1 \over 4\pi} \sum\limits_{k=1}^{p-1}~{\rm ln}~\bigg[r^{2} + r'^{2} - 2rr'
{\rm cos}~  \bigg(
\D \theta + {2\pi k \over p}\bigg)\bigg]\cr}\eqno(2.18)$$

\n i.e., a test charge
at $(r, \theta)$ {\it sees} $~p-1$ images of itself located at $(r, \theta +
 {2 \pi k \over p})$ ($k = 1, 2, \ldots p-1$.)

\n The proof of this assertion is straightforward and is given in appendix A.
(The method of images is illustrated in Figure 1 for $p = 4$).

\bigskip

\baselineskip 8 pt
\section{3. Vacuum polarisation near a point mass in 2+1}

\ssection{~~~~Dimensions}

\baselineskip 18 pt

\n It is well known that in a large variety of situations, the imposition
of nontrivial boundary condition serves to alter the zero point
fluctuations of a quantum field leading to the existence of a vacuum
polarisation (DeWitt$^{13}$). One might conjecture that the conical boundary
conditions
implicit in metric (1.5) will also lead to similar effects arising in the
space-time of a massive point particle in 2+1 dimensions.

To investigate this possibility we shall consider a massless scalar field $\phi
(x)$
propagating in the conical background geometry (1.5), and satisfying the field
equation

$$\dal~ \phi (x) + \xi R = 0\eqno(3.1)$$

\n (where the curvature scalar $R$ is taken to be equal to zero everywhere
except
 at the location of the point mass; $\xi = {1\over 8}$ corresponds
to conformal coupling in 2+1 dimensions$^{14}$). Using
conventional canonical quantisation techniques, the field operator $\phi (x)$
 may be expanded as a mode sum

$$\eqalign{\phi (x) = &\sum_\l \bigg( a_\l u_\l (x) + a_\l^\dagger u^{*}_\l (x)
\bigg),
\cr&x \equiv (t,r,\t)\cr}\eqno(3.2 )$$

\n where $a_\l , a_\l^\dagger$ are annihilation and creation operators,
satisfying the
commutation relations $[ a_\l , a_{\l'} ] = \d_{\l\l'}$. The mode functions
$u_\l (x)$ satisfy the differential equation (for $r> 0$)

$$\bigg( {\partial^2 \over \partial t^2} - {1\over r} {\partial \over \partial
r} r
{\partial \over \partial r} - {1\over r^2} {\partial^2 \over \partial \t^2}
\bigg) u_\l (x) = 0\eqno(3.3a)$$

\n and the boundary conditions

$$u_\l (r, \t) {\biggl |_{r = R} } = 0\eqno(3.3b)$$

$$u_\l (r,\t) = u_\l \bigg (r, \t + {2\pi \over p}\bigg) \eqno(3.3c)$$

\n The essential features of the conical spatial geometry are incorporated in
the
boundary condition (3.3c). The boundary condition (3.3b) is imposed to
facilitate
normalisation and mode counting. We shall take the $R\rightarrow \infty$ limit
at
a convenient point, later on in our discussion. Equation (3.3a) can be solved
exactly and its solution expressed as

$$u_\l \equiv u_{lm} (r,\t, t) = N_{lm} J_{p\vert m \vert} (\omega_l r)
e^{ipm\t}
e^{-i\omega_l t}\eqno(3.4)$$

\n where $\omega_l = \xi_l /R,~~\xi_l$ being the $l^{\rm th}$ zero of
$J_{p\vert m
\vert}
(x)$, $m = 0, \pm 1, \ldots , ~~~l = 1,2,\ldots$

\n $N_{lm}$ is a normalisation constant whose value is fixed using the
canonical
equal time commutation relation

$$[ \phi (\vec x , t), \pi (\vec x~' , t)] = i \d^2
(x- x')\eqno(3.5)$$

\n where the conjugate momentum $\pi (\vec x , t) = \dot \phi (\vec x, t)$.
(3.5)
is equivalent to the condition

$$\int d^2 x \vert u_{lm} \vert^2 = (2 \o_l)^{-1}\eqno(3.6)$$

\n which yields a normalisation constant $N_{lm}$ given by

$$N_{lm} = \bigg({p\over 2\pi \o_l R^2}\bigg)^{1/2} \bigg[ J_{p\vert m \vert +
1}
(\xi_l )\bigg]^{-1}.\eqno(3.7)$$

\n The annihilation operator $a_\l$ defined in (3.2) defines a vacuum $\vert
0\rangle
 ~( a_\l \vert 0 \rangle = 0)$ in which the two point (Wightman) function $D_p
(x, x')$
can be expressed as a mode-sum$^{14}$

$$\eqalign{D_p (x, x') &=\langle \phi (x) \phi (x') \rangle = \sum_\l u_\l (x)
u_\l^{*} (x')\cr
&= \sum\limits^\infty_{l=1} \sum\limits^\infty_{m=-\infty} N^2_{lm} u_{lm} (x)
 u^{*}_{lm}(x')\cr}\eqno(3.8)$$

\n The dummy boundary $r = R$ is removed at this stage by taking $R \rightarrow
\infty$ in (3.8) and by noting that

$$\eqalign{\lim_{R\rightarrow \infty}&{1\over R} \sum\limits^\infty_{l=1}
u_l \rightarrow {1\over \pi} \int\limits^\infty_0 d\o ~u_l \cr
\lim_{R\rightarrow \infty} &J^2_{p\vert m \vert + 1} (\xi_l ) \rightarrow
{2 \over \pi \xi_l}
= {2\over \pi \o_l R}~~.\cr}\eqno(3.9)$$

\n The two point function is obtained as an integral over $\omega$ and a
summation
over $m$

$$D_p (x, x') = {p\over 4\pi} \sum\limits^\infty_{m=-\infty} e^{imp(\t - \t')}
\int\limits^\infty_0 d\o e^{-i\o (t - t')} J_{p\vert m \vert} (\o r)
J_{p\vert m \vert} (\o r').\eqno(3.10)$$

\n Carrying out the integration over $\omega$ and setting

$${\rm cosh~}u_0 = {r^2 + r'^2 - (t - t')^2 \over 2rr'}\eqno(3.11a)$$

\n we obtain$^{12}$

$$D_p (x, x') = {p \over 4\pi^2} {1\over (2rr')^{1/2}} \int\limits^\infty_{u_0}
 {du \over ({\rm cosh}~ u - ~{\rm cosh}~u_0)^{1/2}}\sum\limits^\infty_{m
=-\infty}
 e^{imp(\t - \t')-\vert m \vert pu}.\eqno(3.11b)$$

\n The summation over $m$ can be obtained in a closed form and the two point
function finally reduces to

$$D_p (x, x') = {p \over 4\pi^2} {1\over (2 rr')^{1/2}}
\int\limits^\infty_{u_0}
{du \over ({\rm cosh}~u - {\rm cosh}~u_0)^{1/2}}
{{\rm sinh}~pu \over ({\rm cosh}~pu - {\rm cos}~p(\t - \t'))}~~.
\eqno(3.12)$$

\n It is straightforward to see that for $p =1$ one recovers the standard
Minkowski space two point function

$$\eqalign{D_1 (x, x')&= {1\over 4\pi^2 (2rr')^{1/2}}\int\limits^\infty_{u_0}
{du \over ({\rm cosh}~u - {\rm cosh}~u_0)^{1/2}}
{{\rm sinh}~u \over ({\rm cosh}~u - {\rm cos}~(\t - \t'))}\cr
&={1\over 4\pi \sigma}~~~~~~~~~~~~~~~~~{\rm where}~~~\sigma =
 \vert x - x'\vert .\cr}\eqno(3.13)$$

\n The two point function $D_p (x, x')$ obtained in (3.12) must be renormalised
by subtracting out the Minkowski space contribution from it$^{13, 14}$, so that

$$\eqalign{D_p (x, x')_{\rm ren} & = D_p (x, x') - D_1 (x, x')\cr
& = {1\over 4\pi^2} {1\over (2rr')^{1/2}}
\int\limits^\infty_{u_0} {du \over ({\rm cosh}~u - {\rm cosh}~u_0 )^{1/2}}
\bigg[ {p~{\rm sinh}~pu \over {\rm cosh}~pu - {\rm cos}~p (\t - \t')} \cr
&~~~~~~~~~~~~~~~~~~~~~~~~~~~~~~-
{{\rm sinh}~u \over {\rm cosh}~u - {\rm cos}~(\t -
\t')}\bigg]~.\cr}\eqno(3.14)$$

\n At this stage one is in a position to evaluate the renormalised vacuum
 expectation values of the zero point fluctuations of the field $\langle
\phi^2 (x)\rangle$ and its energy-momentum tensor
$\langle T_\nu^\mu (x)\rangle$.

\n Given the propagator on an arbitrary 2+1 dimensional manifold, the vacuum
energy
momentum tensor may be determined by$^{14}$

$$\eqalign{\langle T_\nu^\m (x) \rangle = \lim_{x' \rightarrow x}
\bigg\{&(1-2\xi)
\bigtriangledown^\mu \bigtriangledown'_\nu - \bigg( {1\over 2} - 2 \xi \bigg)
g^\m_\nu \bigtriangledown_\l \bigtriangledown'^\l\cr
& - 2\xi \bigtriangledown^\m \bigtriangledown_\nu
+ {2 \over 3} \xi g^\m_\nu \bigtriangledown_\l \bigtriangledown^\l
-\xi \bigg[ R^\m_\nu
-{1\over 2} R g^\m_\nu + {4\over 3} \xi Rg^\m_\nu \bigg]\cr
&+ \bigg( {1\over 2} - {4\over 3} \xi \bigg) m^2 g^\m_\nu \bigg\}
 D_p (x, x')_{\rm ren}~~.\cr}
\eqno(3.15)$$

\n For a massless field in flat spacetime (3.15) reduces to

$$\langle T^\m_\nu (x) \rangle = \lim_{x' \rightarrow x} \bigg[ (1-2\xi)
g^{\m\l} \partial_\l \partial'_\nu - \bigg({1\over 2} - 2\xi \bigg) g^\m_\nu
g^{\l\a} \partial_\a\partial'_\l - 2\xi g^{\m\l} \bigtriangledown_\l
\partial_\nu \bigg] D_p (x, x')_{\rm ren}~.\eqno(3.16)$$

\n The coincidence limits of the various derivatives involved in (3.16) can all
be related to the two quantities
$\lim_{\theta' \rightarrow \theta} {\partial^2 \over \partial \theta^2}
 D_p (\theta, \theta')_{\rm ren}$ and $\lim_{\t' \rightarrow \t} D_p (\t,
\t')_{\rm ren}$ in the following manner$^{15}$

$$\lim_{\t' \rightarrow \t} {1\over r^2} {\partial^2 \over \partial \t \partial
\t'}
 D_p (\t, \t')_{\rm ren} = \lim_{\t' \rightarrow \t} - {1\over r^2}
{\partial^2 \over \partial \t \partial \t} D_p (\t, \t')_{\rm ren}$$

$$\lim_{x' \rightarrow x} {\partial^2 \over \partial r \partial r'}
D_p (x, x')_{\rm ren}= \lim_{\t' \rightarrow \t} {1\over 2r^2} \bigg( {3\over
4} + {\partial^2 \over \partial \t^2}\bigg) D_p (\t, \t' )_{\rm ren}$$

$$\lim_{x'\rightarrow x} {\partial^2 \over \partial r^2} D_p (x, x')_{\rm ren}
=
\lim_{\t' \rightarrow \t} {1\over 2r^2} \bigg({5\over 4} - {\partial^2 \over
 \partial \t^2} \bigg) D_p(\t, \t')_{\rm ren}$$

$$\lim_{x' \rightarrow x} {\p^2 \over \p t^2 } D_p (x, x')_{\rm ren} =
\lim_{x' \rightarrow x} - {\p^2 \over \p t \p t'} D_p (x, x')_{\rm ren} =
\lim_{\t' \rightarrow \t} {1\over 2r^2} \bigg({1\over 4} + {\p^2 \over \p
\t^2}\bigg)
D_p (\t, \t')_{\rm ren}$$

$${\rm and,~~using}~~~\dal ~ D(x, x')_{\rm ren} = 0~,$$

$$\lim_{x' \rightarrow x} {1\over r} {\p \over\p r} D_p (x, x')_{\rm ren} =
\lim_{\t' \rightarrow \t} - {1\over 2r^2} D_p (\t, \t')_{\rm ren}.\eqno(3.17)$$

\n The quantities $\lim_{\t' \rightarrow \t} D_p (\t, \t')_{\rm ren}$ and
$\lim_{\t' \rightarrow \t} {\partial^2 \over \partial \t^2} D_p (\t, \t')_{\rm
ren}$
can be expressed in terms of finite integrals

$$\eqalign{\lim_{\t' \rightarrow \t} D_p(\t, \t')_{\rm ren} = \langle \phi^2
(r) \rangle&=
{1\over 8\pi^2 r} \int\limits^\infty_0 {du \over {\rm sinh}~u} [p~{\rm coth}~u
- ~{\rm coth}~u]\cr
&\equiv {1\over 8\pi r} s_1 (p)\cr}\eqno(3.18)$$

$$\eqalign{\lim_{\t' \rightarrow \t} {\p^2 \over \p \t^2} D_p (\t, \t')_{\rm
ren}
&= {1\over 8 \pi^2 r} \int\limits^\infty_0 {du \over {\rm sinh}~u} \bigg[ {{\rm
coth}~u \over {\rm sinh}^2 u}
- {p^3 {\rm coth}~pu \over {\rm sinh}^2 pu }\bigg] \cr
&\equiv {1\over 16 \pi r} s(p)\cr}\eqno(3.19)$$

\n Substituting the relations (3.17) in (3.16) one obtains

$$\eqalignno{\langle T^\m_\nu (x) \rangle =  &{1\over 2r^2} \bigg[ \lim_{\t '
\rightarrow \t} {\partial^2 \over \partial \t^2} D_p (\t, \t')_{\rm ren}~{\rm
diag}~[-1,-1,2]\cr
&+ \bigg( 2\xi - {1\over 4} \bigg) \lim_{\t' \rightarrow \t}
D_p (\t, \t')_{\rm ren}~{\rm diag}~[-1,1,-2] \bigg]\cr
& ~~~\cr
&= {1\over 32 \pi r^3}  \bigg[ s(p) ~{\rm diag}~[-1,-1,2]\cr
&+  (4\xi - {1\over 2}) s_1 (p)  ~{\rm diag}~[-1,1,-2] \bigg].&(3.20)\cr}$$

\n $s_1(p) = 8\pi r \langle \phi^2\rangle~$ and $s(p) = 32 \pi r^3 \langle
 T_{00} \rangle_{ \xi = {1 \over 8}}$ are shown plotted against $p$ in Figures
2 and 3.

\n The vacuum expectation value of the energy momentum tensor so obtained
satisfies the
conservation equations $\langle T^\m_\nu (x) \rangle_{;\m} = 0$ and is
traceless for
$\xi = 1/8$.

\n It is interesting to note that as in the case of the electrostatic field,
the
method of images can once more be used for integer values of $p$, to evaluate
the
two point function $D_p (x, x')$. Using this approach which is outlined in
Appendix A, we find

$$D_p (x, x')_{\rm ren} = \sum\limits^{p-1}_{k=1} D_{\rm Mink} (x, x'_k
)\eqno(3.21)$$

\n where $ x \equiv (r, \t, t),~~~~~ x~' \equiv  (r', \t' + {2\pi k \over p},
t')$.

\n Using (3.16) we find that $\langle T^\m_\nu (x) \rangle$ has precisely the
same
form as (3.20) with $s_1 (p)$ and $s(p)$ now being the finite sums

$$\eqalign{s_1(p) & = \sum\limits^{p-1}_{k=1} {\rm cosec} {\pi k \over p}\cr
s (p) & = \sum\limits^{p-1}_{k=1}  \bigg({\rm cosec}^3 {\pi k \over p} -
{1\over 2}
{\rm cosec}
{\pi k \over p}\bigg)\cr}\eqno(3.22)$$

\n The integrals (3.18 ) and (3.19) do reduce to the sums (3.22) for integer
values of $p$ as can be verified by means of
contour integration (see Appendix C).

\n It is also interesting to evaluate the total vacuum energy associated with a
 localised object of mass $M$

$${\cal E} = \int\limits_0^{2\pi} \int\limits^\infty_{M^{-1}} \langle T_{00}
(r)
\rangle rdrd\t \eqno(3.23a)$$

\n (where the lower limit $M^{-1}$ has been imposed in order to make ${\cal E}$
a
 finite quantity) as a result

$${\cal E} = {-M \over 32 \pi} \bigg[ s(p) + (4\xi - {1\over 2} ) s_1 (p)
\bigg]
\eqno(3.23b)$$

\n where $p = (1 - 4 G_2 M)^{-1}.$

\n We find that for $\xi \geq 0$, ${\cal E} < 0$; a consequence of the fact
that for scalar fields with $\xi \geq 0$ the energy density associated with
the vacuum polarisation $\langle T_{00}\rangle$, is {\it always} negative.

\bigskip


\section{4. Semi-Classical Einstein Gravity in 2+1 dimensions}

Using the regularised vacuum expectation value for the energy momentum tensor,
 obtained in the previous section,
we can attempt to solve the semi-classical Einstein equations

$$G_{\mn} = \kappa  \langle T_{\mn} \rangle,~~~
\bigg( \kappa = {8\pi G_2 \over c^4}\bigg)\eqno(4.1)$$

\n at a linearised level, in order to obtain the first order metric
perturbation
associated with the back reaction of the vacuum polarisation $\langle T_{\mn}
 \rangle$, on the space-time geometry$^{16, 17}$.We shall look for
self-consistent
static solutions to (4.1).

\n As demonstrated in the previous section $\langle
T^\nu_\mu \rangle$ has the form

$$\kappa \langle T_\m^\nu (r) \rangle =  {A  \over r^3} ~{\rm diag}~[-1,-1,2] +
 {B \over r^3}~{\rm diag}~[-1,1,-2]\eqno(4.2)$$

\n with $A = {l_p\over 32 \pi} s(p), ~~B = {l_p\over 32 \pi} (4 \xi - {1\over
2})
s_1 (p)$~$(l_p = 8\pi G_2 {\hbar \over c^3}$ is the planck length in 2+1
dimensions).
Since $\kappa \langle T_\m^\nu (r) \rangle $ is a function of $r$ alone, one
would
expect the geometry of the perturbed
metric  to respect axial symmetry. The most general form for such a metric
is$^1$

$$ds^2 = e^{2\Phi (r)} (dt^2 - dr^2) - e^{2 \Psi(r)} d\t^2.\eqno(4.3)$$

\n In the perturbative approach which we adopt, we shall expand the metric
about
the flat space solution

$$ds^2 = dt^2 - dr^2 - \bigg( {r\over p} \bigg)^2 d\t^2 \eqno(4.4)$$

\n so that each of $\Phi (r)$ and $\Psi(r)$ may be written as

$$\eqalign{\Phi (r) & = \Phi_c + \phi(r) = \phi(r)\cr
\Psi (r) & = \Psi_c (r) + \psi (r) = {\rm ln} {r\over p} + \psi (r)
\cr}\eqno(4.5)$$

\n where $\Phi_c = 1,$ and $\Psi_c (r) = {\rm ln}~{r\over p}$ are the lowest
order
 terms corresponding to the classical metric
(4.4), $\phi(r)$ and $\psi(r)$ are the first order
corrections in the planck length $l_p$ .

\n The Einstein equations (4.1), with $\langle T_\mn \rangle$ given by (4.2),
when
linearised in $\phi(r)$ and $\psi(r)$ yield

$${d^2 \psi \over dr^2} - {1\over r} {d\phi \over dr} + {2\over r} {d\psi \over
dr}
= {2\pi \over r^3} (A+B)\eqno(4.6a)$$

$${1\over r} {d\phi \over dr} = {2\pi \over r^3} (A-B)\eqno(4.6b)$$

$${d^2 \phi \over dr^2} = -{4\pi\over r^3} (A-B)\eqno(4.6c)$$

\n (We shall now adopt the natural units $G_2 = c = \hbar = 1$, consequently,
 all length scales
will be measured in units of $l_p$ --- the Planck length in 2+1 dimensions.)

\n The functions $\phi (r)$ and $\psi(r)$ can be obtained by integrating
(4.6), giving

$$\phi(r) = {-2\pi \over r} (A - B) + k_1\eqno(4.7a)$$

$$\psi (r) = {-4\pi A\over r} {\rm ln}~ (r + 1 ) + {k_2\over r} +
k_3\eqno(4.7b)$$

In the above equations the constants of integration $k_1$ and $k_3$ must be set
to zero
since it is not possible to have them linear in $l_p$ and dimensionless too.
$k_2$ can also be set to zero since it reflects a scaling $r \rightarrow \alpha
r$.

The line element of the metric (4.3) to first order in $l_p$ now reads

$$ds^2 = \bigg[ 1- {2\pi \over r} (A - B) \bigg] [dt^2 - dr^2] - {r^2 \over
p^2}
\bigg[ 1 - {4\pi A\over r }{\rm ln} r \bigg]d \t^2\eqno(4.8)$$

\n The above approximation to the metric is valid so long as first order
corrections
are small, i.e., when both ${2\pi (A-B) \over r}$ and ${4\pi A\over r}$ are
small
compared to unity.

\n Let us define a new radial coordinate $R(r)$ such that

$$dR = \bigg( 1 - {2\pi \over r} (A-B) \bigg)^{1/2} dr \simeq \bigg( 1 - {\pi
\over r} (A-B)\bigg) dr\eqno(4.9)$$

\n then at large distances from the point
mass $(R \gg 1)$ the line element (4.8) can be rewritten as

$$ds^2 = \bigg[ 1 - {2\pi (A-B) \over R} \bigg] dt^2 - dR^2 - {R^2 \over p^2}
\bigg[ 1 - {2\pi (A+B)\over R} {\rm ln} R \bigg] d\t^2 \eqno(4.10)$$

\n One finds that although the first order metric (4.10) is no longer locally
flat,
its $(R-\t)$ section is still conical, the deficit angle now depending upon
$R$ --- the proper radial distance from the point mass. To quantitatively
describe
this behaviour it is convenient to introduce the deficit angle $\D\t = 2\pi -
{C\over R}$, where $C$ is the circumference of a circle centered around the
point
mass at a fixed proper radius $R$ from it. Then, for the metric (4.10)

$$\D \t (R) = 2\pi \bigg[ 1 - {1\over p} \bigg( 1 - {2\pi (A+B) \over R} {\rm
ln}
 R\bigg)\bigg]\eqno(4.11)$$

\n or, in terms of the classical deficit angle $\D \t_{class} = 2\pi - {2\pi
\over p}$,

$$\D \t(R) = \D\t_{class} + {2\pi \over p} {(A+B)\over R} {\rm ln} R
\eqno(4.12)$$

\n One finds that for negative values of the energy density $(A+B > 0)$, the
deficit
angle {\it increases} as the point mass is approached. For positive values of
the
 energy density however, the deficit angle is seen to {\it decrease} with $R$.

\n At large distances from the point mass
$\Delta \theta (R) \simeq \Delta \theta_{\rm class}$, and
the local geometry of the space-time approaches the asymptotic form described
by
the classical metric (4.4).

\n An important consequence of the linearised metric (4.10) is the existence of
a well
defined Newtonian limit to the semi-classical Einstein's equations (4.1) in 2+1
dimensions.

\n A given space-time geometry is usually said to admit a Newtonian limit if
the time-time component of its metric tensor has the form$^{18}$

$$g_{00} = \bigg(1+ {2\Phi (r) \over c^2}\bigg)\eqno(4.13)$$

\n $\Phi (r)$ then plays the role of the Newtonian potential and, in the slow
motion
limit, the acceleration of a test particle is determined by
${d^2 \vec x \over dt^2} = - \bigtriangledown \Phi$. The Einstein equations in
the
same limit assume the form

$$ R^0_0 = \D \Phi = 4\pi G (T_0^0 - {1\over 2} T)\eqno(4.14)$$

\n ($T^\nu_\m \rightarrow \langle T_\m^\nu \rangle$ in our case)

\n From (4.10) and (4.13) we find that

$$\Phi (R) =- G_3{\pi (A-B)\over R}\eqno(4.15)$$

\n where
$G_3~ (\equiv G_2 l_{pl})$ is the Newtonian gravitational constant
 {\it in 3+1 dimensions.}

\n For a conformally coupled field $B=0$ and $\Phi(R) = - { G_3 \pi  A \over
R}$.
A plot of $A$ against $M$ --- the mass of the point particle(Figure (4)) shows
that in a broad range of parameter space,
$A $ is approximately proportional to $M$.

\n The above results prompt us to define a {\it gravitating} mass $M_G$

$$M_G = \pi (A-B)~= {m_{pl} \over 32} \bigg[ s(p) - (4 \xi - {1\over 2}) s_1
(p)
\bigg] \eqno(4.16)$$

\n where $p = (1-4 G_2M)^{-1}$ so that $\Phi = G_3{M_G \over R}$ ($M_G > 0$ for
${1\over 8} \geq \xi > 0).$

\n Most of the results of the preceding analysis can be easily extended to
other
 massless conformally coupled fields such as massless spinors and vectors. The
conservation equation $\langle T^\m_\nu \rangle ;_\m = 0$ and the trace-free
condition
$\langle T^\m_\m \rangle = 0$, in this case guarantee that the regularised
vacuum
expectation value of the energy momentum tensor have the form

$$\langle T_\m^\nu \rangle = {A_\alpha (p) \over r^3} ~{\rm
diag}~[-1,-1,2].\eqno(4.18)$$

\n The numerical value of the constant $A_\alpha$ depends upon the contribution
to
the vacuum polarisation from a quantum field having spin $\alpha$. For
conformally
 coupled scalars $A_0 = {s(p) \over 32 \pi}$, as demonstrated in the previous
section.
In general, when considering the net contribution to the vacuum polarisation
from
fields with different spin, one would expect $A_\alpha (p)$ to be replaced by
$\bar A(p) = \sum\limits_\alpha n_\alpha A_\alpha (p)$, $n_\alpha$ being the
number
 of spin $\alpha$ fields present in
nature. The Newtonian potential will then assume the somewhat more general\
form
$\Phi (R) = -{\pi G_3 \bar A (p) / R}$.

\n We should note that, strictly speaking, the potential $\phi \sim R^{-1}$
corresponds
to the Newtonian potential in 3+1 dimensions and not in 2+1 dimensions where
the Newtonian potential has the form $\phi \sim ln r$. Thus a test particle
near a
point mass in 2+1 dimension behaves just as if it were in the neighborhood of a
 gravitating mass in 3+1 dimension with its
motion being confined to a 2D section of the 3D space.

\n The above results acquire a special significance in view of the fact that
general relativity does not posses a Newtonian limit in 2+1 dimensions. (A
consequence
of the lack of propagating modes in Einstein gravity in dimensions lower than
4.)
The attempt to construct alternate theories of gravitation which might
have a well defined Newtonian limit in lower dimensions has also proved to be
very
 elusive$^{2,3,4,19}$. For instance the Einstein--Cotton equations (1.4),
describing
topologically massive gravity, do endow the gravitational field with a
nontrivial
dynamics but not with a Newtonian limit.

\bigskip


\baselineskip 8 pt

\ssection{5. Quantum effects in the 2+1 dimensional}

\ssection{ ~~~~de Sitter - Schwarzschild metric}

\baselineskip 18 pt

\n In this section we extend our previous analysis to non-static conical
 space-times such as the de Sitter-Schwarzschild metric which describes
 the space-time of a point mass in the presence of a homogeneous
cosmological constant $\Lambda$. The line element for this space-time has the
form

$$ds^2 = dt^2 - e^{2Ht} (dr^2 + {r^2 \over p^2} d\t^2)\eqno(5.1)$$

\n where $p = (1-4 M)^{-1},~~~~0\leq \t < 2\pi$ and
$ ~H = \sqrt{{\Lambda \over 3}} .$

\n For integer values of $p$ the two point function in this space can be
expressed as
a finite sum over the Green's function in de Sitter space $G (x, x')$, using
the
 method of images described in Appendix A. As a result we get

$$G_p (\vec x, \vec x~') = \sum\limits^{p-1}_{k=0} G (\vec x, \vec
x~'_k)\eqno(5.2)$$

\n where $x \equiv (r,t,\eta)$ and $x'_k \equiv (r' , \t' + {2\pi k \over p} ,
\eta')$,
$\eta$ being the conformal time coordinate $\eta = \int {\rm dt}~ e^{-Ht}$.

\n The de Sitter space propagator $G (\vec x, \vec x~')$ has been obtained in
Appendix B for the general case of an n-dimensional space-time. In our
treatment we
shall set $n = 3$ and shall regard the mass of the scalar field to be small
($m/H < 1$).
In terms of the conformal time, the scale factor assumes the form
$a = -{1\over H\eta}$,
consequently the proper distance to the point mass in (5.1) is given by
$R = a r = {-r \over H\eta} $.

\n The $k=0$ term in (5.2) when suitably regularised and differentiated
gives the one-loop vacuum
energy-momentum tensor in 2+1 dimensional de Sitter space. The additional
contribution to the de Sitter propagator (B.19) due to the conical nature of
the space-time (5.1) is given by

$$G_p (x, x')_{\rm cone} = {-H\over 4\pi} \nu {~\rm cosec}~\pi \nu~
\sum\limits^{p-1}_{k=1} F ( 1+ \nu , 1 - \nu ; {3\over 2} ; \o_k )\eqno(5.3)$$

\n where

$$\nu = \bigg[ 1 - {m^2 \over H^2} - 6 \xi\bigg]^{1/2}$$

\n and

$$\o_k = 1 -{r^2 + r'^2 - 2rr' {\rm cos} (\t - \t' + {2\pi k \over p}) - \D
\eta^2
\over 4\eta \eta '}.$$

\n In obtaining (5.3) we have substituted the value of the de Sitter propagator
 $(B.19)$ into (5.2). Just as in $\S$3, we compute the expectation values
$\langle \phi^2 (x) \rangle$ and $\langle T^\m_\nu (x) \rangle$ from the
coincidence
limits of $G_p (x, x')$ and its various derivatives. As a result

$$\eqalign{\langle \phi^2 (x) \rangle_{\rm cone} &=\lim_{x' \rightarrow x}
G_p (x, x')_{\rm cone}\cr
&= {-H\over 4\pi} \nu {~\rm cosec}~\pi \nu~ \sum\limits^{p-1}_{k=1}
F\bigg(1 + \nu, 1- \nu;
{3 \over 2}; 1 - {r^2  {\rm sin}^2 {\pi k \over p}\over
\eta^2}\bigg)\cr}\eqno(5.4a)$$

\n For conformally coupled massless scalar fields $\nu = {1\over 2}$,and we
find that
$ \langle \phi^2 (x) \rangle_{\rm cone}$ is conformally related to the flat
spacetime
result obtained in (3.18):

$$\langle \phi^2 (x) \rangle_{\rm cone} = -H \eta \langle \phi^2 (x)
\rangle_{{\rm flat}}  = {H  \over 8\pi R} s_1 (p)\eqno(5.4b)$$

\n where $s_1 (p) = \sum\limits^{p-1}_{k=1} ~{\rm cosec}~{\pi k \over p}.$

\n The vacuum expectation value of the energy-momentum tensor $\langle T^\m_\nu
(x)
\rangle$ is
obtained by means of the general relationship (3.16). After some lengthy
calculations
 we obtain an expression for $\langle T^\m_\nu (x) \rangle$ in terms of the
coincidence limit of
the hypergeometric function $F(1 - \nu, 1+ \nu; 3/2; \o_k )$ and its first and
second
derivatives with respect to $\o_k$ .
The derivatives of a hypergeometric function may be evaluated using$^{20}$

$${d \over dz} F (a,b; c; z) = \bigg({ab \over c}\bigg) F (a+1, b+1; c+1; z).$$

\n We find that for the case of a conformally coupled massless scalar field,
 $\langle
T^\m_\nu \rangle$ is conformally related to the flat spacetime result.

$$\langle T^\m_\nu (x) \rangle_{\rm cone} = (-H\eta)^3 \langle T^\m_\nu (x)
 \rangle_{{\rm flat}}.
\eqno(5.5)$$

\n with $\langle T^\mu_\nu (x) \rangle_{{\rm flat}}$ given in (3.20).

\n For massless conformal fields $\langle T^\m_\nu (x) \rangle_{\rm cone}$
as evaluated above provides the {\it entire} contribution to the vacuum
 expectation value of the energy
momentum tensor in the de Sitter -- Schwarzschild metric. This is due to the
fact
that the $k = 0$ term in (5.2) (which we had dropped while calculating
$\langle T^\m_\nu (x) \rangle_{\rm cone}$) when suitably differentiated and
regularised gives the one loop vacuum expectation value of the energy momentum
tensor in 2+1 dimensional de Sitter space. Since the only maximally form
 invariant rank two tensor
under the de Sitter group is $g_\mn$, the entire vacuum energy momentum tensor
in de Sitter space can be constructed out of its trace~: $\langle T_\mn (x)
\rangle = g_\mn {\langle T \rangle \over n}$ ($n$ being the dimensionality
of the space-time). For massless conformally coupled fields in odd dimensions
$\langle T \rangle = 0$ so that $\langle T^\m_\nu (x) \rangle = 0$ in any odd
dimensional de Sitter space [Birrell and Davies$^{14}$ p.177, p.191].
Consequently, while evaluating the energy momentum tensor for conformally
coupled fields the $k = 0$ term in (5.2) will not contribute and
$ \langle T^\m_\nu (x) \rangle_{\rm ren} = \langle T^\m_\nu (x) \rangle_{\rm
cone}$.

\n For general values of $m$ and $\xi$ ,$\langle \phi^2 (x) \rangle$ and
$\langle
T^\m_\nu (x) \rangle$ can be expressed in terms of
elementary functions in the assymptotic regimes $R \ll H^{-1}$ and $R \gg
H^{-1}$
(equivalently ${r\over \eta} \ll 1$ and ${r\over \eta} \gg 1$) using the
well known linear transformation formulae for the hypergeometric
function$^{20}$.

\n As a result we find (for $R \ll H^{-1}$)

$$\langle \phi^2 (x) \rangle_{\rm cone} = {1 \over 4\pi R} s_1 (p).
\eqno(5.6)$$

\n The diagonal components of $\langle T^\m_\nu (x) \rangle_{\rm cone}$ assume
the form

$$\eqalign{\langle T^\m_\nu (x) \rangle_{\rm cone}^{\rm diag} = &{1 \over 32
\pi R^3} \bigg[ s(p) ~{\rm diag}~
[-1,-1,2] \cr
&+ \bigg(4 \xi - {1\over 2} \bigg) s_1 (p)~{\rm diag}~[-1,1,-2]\bigg]
\cr}\eqno(5.7)$$

\n where $s_1(p)$ and $s(p)$ are defined in (3.18) and (3.19).

\n One immediately finds that $\langle \phi^2 (x)\rangle_{\rm cone}$ and
$\langle T^\m_\nu (x) \rangle_{\rm cone}^{\rm diag}$
obtained in (5.6, 5.7) are conformally related to the flat spacetime results
(3.18)
and
(3.20) {\it for all values of $\xi$}. Both $\langle \phi^2 \rangle_{\rm cone}$
and $\langle T^\m_\nu \rangle_{\rm cone}^{\rm diag}$ also seem to be
independent
of mass $m$ in this limit.
In addition to $\langle T^\m_\nu (x)\rangle_{\rm cone}^{\rm diag}$
given by (5.7) there also exists an energy flux given by
the off diagonal component

$$\langle T^\eta_r \rangle_{\rm cone} = {H \over 16 \pi R^2} \bigg( 4 \xi -
{1\over 2 } \bigg) s_1 (p). \eqno(5.8)$$

\n The energy flux is a measure of the energy flowing away from the point
source
(it is absent in the case of a homogeneous and isotropic space-time such as the
 2+1 dimensional de Sitter metric). One can see from (5.8) that the direction
of
the energy flux is {\it outwards} from the origin for $\xi > {1\over 8}$,and
{\it inwards} for $\xi < {1 \over 8} $

\n It may be noted that the diagonal components $\langle T^\m_\nu (x)
\rangle_{\rm cone}^{\rm diag}$ are of one order higher in $(HR)$
than $\langle T^\eta_r (x) \rangle_{\rm cone}$.

\n (We feel that the results (5.6-8) will also remain valid for non integer $p$
if
$s(p)$ and $s_1(p)$ are expressed as integrals using (3.21)).

\n In the other limiting case $R \gg H^{-1}$, $\langle T^\m_\nu (x)
\rangle_{\rm cone}$
can be obtained using the following transformation property of the
hypergeometric
function$^{20}$

$$\eqalign{F(a,b; c; \o ) = &(1-\o)^{-a} {\G (c) \G (b-a) \over
\G(b) \G(c-a)} F\bigg(a,c-b; a-b + 1; {1\over 1-z}\bigg)\cr
&+(1-\o)^{-b} {\G(c) \G(a-b)\over \G(a) \G(c-b)} F\bigg( b, c-a; b-a+1; {1\over
1-z}
\bigg).\cr}\eqno(5.9)$$

\n We give below only the result for
$\langle \phi^2 \rangle_{\rm cone}$ and $\langle T^\eta_r \rangle_{\rm cone}$
to
leading orders in $(HR)$

$$\langle \phi^2 (x) \rangle_{\rm cone} = {H^{2\nu-1} \over 16 \pi} {2 ^{2\nu}
 \over {\rm sin}~\pi \nu}
R^{2\nu-2} \sum\limits^{p-1}_{k=1} {\rm sin}^{2\nu -2}
{k\pi \over p}\eqno(5.10)$$

$$\eqalign{\langle T^\eta_r (x) \rangle_{\rm cone} = {H^{2\nu} \over 16\pi}
 {2^{2\nu} (1-\nu)\over{\rm sin}~
\pi \nu} R^{2\nu - 3} &\bigg[ (6 \xi -1 ) -\nu
(4\xi -1 ) \bigg]  \sum\limits^{p-1}_{k=1} {\rm sin}^{2\nu -2}~ {\pi k \over
p}\cr
}\eqno(5.11)$$

The expressions in
(5.10) and (5.11) are well defined only for $\nu < 1$, for $\nu = 1,2,\ldots$.
(5.10, 5.11) display
an infrared divergence which is similar to the infrared divergence found for
the
massless
propagator in de Sitter space$^{21,22}$.It may be noted that for
$\nu < 1 ~,\langle T_r^\eta \rangle \rightarrow 0 $
as $ R \rightarrow \infty $ so that there is no radiation flux at infinity.

\n We would finally like to point out that although as mentioned earlier in
this section,
vacuum polarisation effects for conformally coupled scalar fields are absent
in a 2+1 dimensional de Sitter space-time, an
Unruh-DeWitt particle detector nevertheless experiences a finite response.

\n To see this we note that
the response function of an  Unruh-DeWitt particle detector is given by$^{23}$

$$F(E) = \int\limits^\infty_{-\infty} d\tau \int\limits^\infty_{-\infty}
d\tau' e^{-iE(\tau - \tau' )} D(x (\tau), x(\tau'))\eqno(5.12)$$

\n where $D(x (\tau), x(\tau'))$ is the positive frequency Wightman Green's
function evaluated for two points lying
on the worldline of the detector.

\n For 2+1 dimensional de Sitter space the response rate for an inertial
 detector is given by

$${F(E)\over T} = \int\limits^\infty_{-\infty} d(\D t) ~~ e^{-iE\D t} D(\D
t)\eqno(5.13)$$

\n where for a massless conformally coupled scalar field (see B.17)

$$D(\D t) = {H \over 16 \pi} {\rm cosec}~{iH\D t \over 2}.\eqno(5.14a)$$

\n It is convenient to rewrite $D(\D t)$ as a sum$^{12}$

$$D(\D t) = {H\over 8 \pi} \bigg[{2 \over iH\D t} + {iH \D t\over \pi^2}
 \sum\limits^\infty_{k=1} {(-1)^k \over \bigg({iH \over 2\pi} \D t + k \bigg)
\bigg( {iH \over 2\pi} \D t - k \bigg)}\bigg]\eqno(5.14b)$$

\n  (5.13) can be evaluated by means of contour integration, so that finally

$${F(E)\over T} = {1 \over 4} ~~{1 \over  e^{2\pi E / H} + 1},\eqno(5.15)$$

\n which describes a thermal Fermi-Dirac distribution at the {\it de Sitter
 temperature}$^{14}$ $T = {H \over 2\pi}$.

\n Equation (5.15) points to a remarkable feature common to all odd-dimensional
space-times ---
that of the inversion of statistics$^{24}$.
This can be viewed to be a consequence of the fact that
the de Sitter Green's function in odd dimensions displays an antiperiodicity in
imaginary time. One can see this in the general case by applying the
 transformation property of the hypergeometric
function$^{20}$

$$F(a,b; c; z) = (1-z)^{c-a-b} F(c-a, c-b; c; z)\eqno(5.16)$$

\n to the n-dimensional Green's function (B.15). As a result we obtain

$$\eqalign{D(t, t') = {\rm cosec}^{n-2} ({iH \over 2} \D t )&~~
{1\over (4\pi)^{n/2}} \bigg({ H \over 2 }\bigg)^{n-2} {\G ( {n-1 \over 2} + \nu
)
 \G ({n-1 \over 2} - \nu)
\over \G  ({n \over 2})}
\cr
&\times F \bigg( {1\over 2} - \nu, {1\over 2} + \nu~; {n \over 2};
{}~ {\rm cosh}^2 {H \D t \over 2}
\bigg)\cr}\eqno(5.17)$$

\n where $\nu = \bigg[ {(n-1)^2 \over 4} - {m^2 \over H^2} - n (n - 1) \xi
\bigg]^{1/2}$.

\n The above expression is manifestly antiperiodic in $\D t$ with period
${2\pi \over H}$
if $n$ is odd, and periodic in $\D t$ with the same period,if $n$ is even.

\n This leads us to conjecture that the response of a particle detector will be
of
the Fermi-Dirac type in odd space-time dimensions even for a massive,
nonconformally
coupled scalar field. (In an even dimensional space-time ${F(E) \over T}
\propto
(e^{{2\pi E\over H}} -1 )^{-1}$ for details regarding the behaviour of particle
detectors in space-times of arbitrary dimension see Takagi$^{24}$.)

\n The response of a particle detector placed
at fixed distance $R$ from the point mass in the conical spacetime (5.1) will
be modified to

$${F(E)\over T} = {H \over 8 \pi} \int\limits^\infty_{-\infty} d(\D t) e^{-iE\D
t}
\sum\limits^{p-1}_{k=0}
{1 \over  \bigg[ (1 - H^2 R^2) {\rm sin}^2 i {H\D t \over 2} + H^2 R^2 {\rm
sin}^2
 {\pi k \over p}\bigg]^{1/2}}\eqno(5.18)$$

\n At distances close to the point mass (i.e., $R \ll H^{-1}$), (5.18) reduces
to

$${F(E)\over T} = {1 \over 4}~~ { p\over e^{2 \pi E/H} + 1} \eqno(5.19)$$

\n which is just the detector response in de Sitter space enhanced by a factor
of
$p$. This result could have been anticipated since the vacuum in de Sitter
space has
now effectively been compressed into a region of space $p$ times smaller, due
to the
 removal of the angular wedge from the space-time.

\bigskip

\vfil\eject

\section{6. Twisted Scalar Fields}

\n As first pointed out by Isham$^{25}$ a twisted variety of scalar and spinor
fields can
be defined on a non simply connected manifold by considering antiperiodic
boundary
conditions along the identified coordinate (see also Ford$^{26}$). We shall
consider
a massless, real, scalar field twisted around the mass point $M$ located at $r
= 0$.
The twisted field $\phi (x)$ obeys the same field equations as (3.1). However,
the
boundary conditions for $\phi$ are now different

$$u_\l (r, \t) = - u_\l (r , \t + {2\pi \over p})\eqno(6.1)$$

Solving the field equation $\dal ~ \phi = 0$ with the new boundary conditions
(6.1) we obtain

$$u_\l (x) = N_{lm} e^{ip(m+{1\over 2}) (\t - \t')} e^{-i\o_l (t-t')}
J_{p \vert m + {1\over 2} \vert} (\o r) J_{p \vert m + {1\over 2} \vert} (\o
r')
\eqno(6.2)$$

\n where $N_{lm}$, the normalisation constant, is identical to the untwisted
case
 (3.7).

\n Using (3.8)  and (3.9) to define the two point function for the twisted
field
$D_p^T (x, x')$ we get

$$D_p^T (x, x') = {p\over 4\pi} \sum\limits^\infty_{m=-\infty} e^{ip(m+{1\over
2})
(\t - \t')} \int\limits^\infty_0 d\o ~e^{-i\o_l (t-t')} J_{p \vert m +
 {1\over 2} \vert} (\o r) J_{p \vert m + {1\over 2} \vert} (\o r') \eqno(6.3)$$

\n The integration over $\o$ is similar to the untwisted case (3.11a) and
making
identical substitutions as in that case we obtain

$$D_p^T (x, x') ={p\over 4\pi^2} {1\over \sqrt{2rr'}} \int\limits^\infty_{u_0}
{du\over ({\rm cosh}~ u -
{\rm cosh}~u_0 )^{1/2}} \sum\limits^\infty_{m = -\infty}
e^{ip(m+{1\over 2}) (\t - \t') - p \vert m + {1\over 2} \vert u }\eqno(6.4)$$

\n The above summation can be rewritten in a closed form

$$\sum\limits^\infty_{m = -\infty} e^{-p \vert m + {1\over 2} \vert u +
 ip ( m + {1\over 2} ) (\t - \t')} = {2 {\rm sinh}~ ( {1\over 2} pu )
{\rm cos}~ {1\over 2} p (\t - \t')\over {\rm cosh}~pu - {\rm cos}~p (\t -
\t')},$$

\n so that the two point function $D^T_P (x, x')$ reduces to

$$D^T_p (x, x') = {p\over 2\pi^2} {1\over \sqrt{2rr'}} \int\limits^\infty_{u_0}
{du  \over ({\rm cosh}~u - {\rm cosh}~ u_0)^{1/2}} { {\rm sinh}~ ( {1\over 2}
pu )
 {\rm cos}~ {1\over 2} p (\t - \t') \over ({\rm cosh}~pu - {\rm cos}~p (\t -
\t'))}
\eqno(6.5)$$

\n We now adopt the same procedure in renormalising $D^T_p (x, x')$ as was used
in
$\S$3, namely, we subtract out the $p =1$ term in (6.5) so that the
renormalised
 two point function becomes

$$\eqalign{D^T_p& (x, x')_{\rm ren} = D^T_p (x, x') - D^T_1 (x, x')\cr
&= {1\over 2\pi^2 \sqrt{2rr'}} \int\limits^\infty_{u_0} {du \over
 ({\rm cosh}~u - ~{\rm cosh}~u_0)^{1/2}}
\bigg[{p ~{\rm sinh}~{1\over 2} pu ~{\rm cos}~{1\over 2} p \D \t \over
 ({\rm cosh}~pu - ~{\rm cos}~p \D \t)}\cr
&~~~~~~~~~~~~~~~~~~~~~~~~~~~~~~~-{{\rm sinh}~{u \over 2} {\rm cos}~
{\D\t \over 2}\over ({\rm cosh}~u - ~{\rm cos}~ \D \t)}\bigg]
\cr}\eqno(6.6)$$

\n As in the case of untwisted fields, the vacuum expectation values
$\langle \phi^2 (x) \rangle^T$ and $\langle T^\m_\nu (x) \rangle^T$ can be
obtained from
the coincidence limit of the renormalised two point function $D^T_p (x, x')$
 and its
second derivative with respect to $\t$. As a result

$$\langle \phi^2 (x) \rangle^T =\lim_{\theta ' \rightarrow \t} D_p^T (\t , \t
')_{\rm ren}
= {1\over 8\pi r} s^T_1 {~(p)}\eqno(6.7a)$$

\n where

$$s^T_1 (p) = {2\over \pi} \int\limits^\infty_0 {du \over {\rm sinh}~u}
[ p ~{\rm cosech}~pu - {\rm cosech}~u ]\eqno(6.7b)$$

\n and

$$\lim_{\t' \rightarrow \t}
{\p^2 \over \p \t^2} D^T_p (\t, \t')_{\rm ren} = {1\over 8\pi r} s^T (p)
\eqno(6.8a)$$

\n where

$$s^T (p) = {2\over \pi} \int\limits^\infty_0 {du \over {\rm sinh}~u}
\bigg[ p^3 \bigg({\rm cosech}^3 pu + {{\rm cosech}~pu \over 2}\bigg) - \bigg(
{\rm cosech}^3 u  + {{\rm cosech}~u \over 2} \bigg)\bigg] \eqno(6.8b)$$

\n Using arguments akin to the ones used in $\S$3 while evaluating $\langle
T_\nu^\mu \rangle$ (see (3.16), (3.17) and (3.20)) we finally obtain

$$\eqalign{\langle T^\m_\nu \rangle^T = &{1\over 2r^2} ~\bigg[ \lim_{\t'
\rightarrow \t}
{\p^2 \over \p \t^2} D^T_p (\t, \t')_{\rm ren} ~{\rm diag}~[-1,-1,2]\cr
&+ ~~\bigg( 2 \xi - {1\over 4}\bigg) \lim_{\t' \rightarrow \t}
D^T_p (\t, \t')_{\rm ren}~{\rm diag}~[-1,1,-2]\bigg] \cr
&~~~~~~\cr
&= {1\over 32 \pi r^3} \bigg[ s^T (p) ~{\rm diag}~
[-1, -1, 2] + ( 4 \xi - {1\over 2} ) s^T_1 (p)\cr
&~~~~~~~{\rm diag}~[-1, 1, -2]\bigg]\cr}\eqno(6.9)$$

\n which is precisely (3.20) with $s(p)$ replaced by $s^T (p)$ and $s_1(p)$
replaced by $s^T_1 (p)$. $s^T_1 (p) = 8\pi r \langle \phi^2\rangle^T$ and
$s^T (p) = 32 \pi r^3 \langle T_{oo}\rangle_{\xi ={1 \over 8}}$
are shown plotted against $p$ in Figure (2 \& 3). It is interesting to note
that
$\langle T_{00}\rangle^T > 0$ for massless conformally and minimally coupled
twisted
fields in contrast to the untwisted case.

\n As in the untwisted case, the vacuum energy-momentum tensor will in general
back
react on the space-time geometry via the semi-classical Einstein equations
 $G_\mn = 8 \pi G_2 \langle T_\mn \rangle^T$, giving rise to the linearised
metric

$$ds^2 = \bigg[1-{2\pi (A^T - B^T)\over R}\bigg] dt^2 - dR^2 -{R^2 \over p^2}
\bigg[
1- {2\pi (A^T+B^T)\over R} {\rm ln}~R \bigg] d\theta^2 \eqno(6.10)$$

\n where $A^T = {l_p \over 32\pi} s^T (p)$ and $B^T = {l_p \over 32 \pi} (4\xi
-1/2)
s^T_1 (p)$ ($l_p$ being the planck length). From Figure (3) we see that
for $0 \leq \xi < {1\over 8}$, $A^T - B^T < 0$ and $A^T + B^T < 0$,so that the
deficit angle in (6.10) {\it decreases}
as the point mass is approached in contrast to the untwisted case.

\n As in the case of untwisted fields the method of images can also be used to
determine
$D^T_p (x, x')$, with $p$ now restricted to even integer values $p = 2,4,6,
\ldots$
(see Appendix A), so that

$$D^T_p (x, x') = \sum\limits^{p-1}_{k=0} (-1)^k D_{\rm Mink} (x,
x'_k)\eqno(6.11)$$

\n Proceeding as in $\S$3 and regularising (6.11) by subtracting out $D^T_1 (x,
x')$
we find that the final form of $\langle \phi^2 (x) \rangle^T$ is given once
more by (6.7a) with $s^T_1 (p)$ now being the finite sum~:
 $s^T_1 (p) = \sum\limits^{p-1}_{k=0} (-1)^k {\rm cosec}~{\pi k \over p}
 + {2\over \pi}$.

\n The method of images can also be applied to obtain the corrections to
 the propagator for twisted
fields in the de Sitter-Schwarzschild metric discussed in $\S$5. Following
the procedure
outlined in Appendix A we find

$$G^T_p (x, x')_{\rm cone} = \sum\limits^{p-1}_{k=1} (-1)^k G (x, x'_k )$$

$$= {-H \over 4\pi} \nu {\rm cosec}~\pi \nu  ~\sum\limits^{p-1}_{k=1} (-1)^k
F (1 + \nu, 1-\nu;
{3\over 2}; 1 - {\D x^2_\m - \D \eta^2 \over 4\eta \eta'})\eqno(6.12)$$

\n where $\D  x^2_k = r^2 + r'^2 - 2rr' {\rm cos}~(\t - \t' + {2\pi k \over
p}),
{}~~H = \sqrt{{\L \over 3}}$ and $\nu = [ 1 - {m^2 \over H^2} - 6 \xi ]^{1/2}.$

\n For a massless, conformally coupled twisted scalar field $\nu = {1\over
2}$,and
$\langle \phi^2 (x) \rangle^T$ will simply be
conformally related to the flat space-time result (6.7) so that

$$\langle \phi^2 (x) \rangle^T_{\rm cone} =  {1 \over 8 \pi R} s^T_1
(p)\eqno(6.13)$$

\n where $ R = -{r \over H\eta}$,

\n this result is also true for light scalars $({m \over H} < 1)$ in the
vicinity
 of the point mass $(R \ll H^{-1})$.

\n In addition, we also find as in the untwisted case, the existence of an
energy
 flux

$$\langle T^\eta_r (x) \rangle^T_{\rm cone} = {H \over 16\pi R^2} (4\xi-{1
\over 2})
s^T_1 (p)\eqno(6.14)$$

which vanishes for  conformally coupled fields$~( \xi ={1 \over 8})$.
\n We notice that $\langle T^\eta_r \rangle_{\rm cone}$ has the opposite sign
to (6.8)
 signifying
that in an expanding universe, if the vacuum flux for untwisted fields is
{\it outwards}, then the corresponding flux for twisted fields is {\it
inwards},
(and vice versa).

\bigskip
\vfil\eject

\section{7. Conclusions \&  Discussion}

\n We have shown how nontrivial boundary conditions can affect the behaviour of
 scalar fields
at both  the classical and quantum levels. At the classical level we find that
the electric field associated with a point charge in a conical space-time is
 distorted, leading the charge to experience a self-force --- in the complete
absence of any other charges in the space-time. At a more fundamental we find
that the point mass --- the source of the conical geometry --- also induces a
 vacuum polarisation in the surrounding space-time which is described by a
finite vacuum expectation value of the energy momentum tensor
$\langle T_\mn\rangle \propto {1 \over r^3} $ ($r$ being the distance to the
point mass).
Quantum effects of a similar nature are also known to arise in the space-time
 of a cosmic string, where the smallness of the string tension~: $G\m <
10^{-6}$
 prevents the effects from becoming large. No such constraint is however
present
in 2+1 dimensions with the result that,
quantum effects can be significant in this case.

\n We have also extended our analysis to an expanding Universe, by considering
the behavior of $\langle \phi^2\rangle$ and  $ \langle T_\mn\rangle $ in a
conical
de Sitter space-time. We find that in this case, in addition to the vacuum
polarisation there also exists a finite vacuum energy flux
$\langle T_{rt}\rangle $ --- describing a net flow of energy  {\it away}
from the point source.

\n As part of our analysis we also evaluate the scalar field propagator in
an  $n$-dimensional
de Sitter space extending previous work by Takagi$^{24}$. We show that if
 $n$ is even , the propagator exhibits a periodicity in imaginary time with
period $ T = {2 \pi \over H} $ ( H being the Hubble constant in
de Sitter space). On the other hand for odd $n$ the propagator displays an
anti-periodicity in imaginary time. This leads us to conclude that the vacuum
 for scalar fields in even/odd dimensional de Sitter space resembles a thermal
bose/fermi distribution
.We are also faced with the surprising result that
 an inertial particle detector registers a finite thermal response in de Sitter
space
of odd dimensions$^{24}$, even though the vacuum expectation value of the
energy
momentum tensor for conformal fields vanishes identically in this case.

\n An important outcome of our analysis is that, at the semi-classical level,
 solutions to the Einstein equations $G_{\mu\nu} = {8\pi G_2 \over c^4}
\langle T_{\mu \nu }\rangle$ possess
a well defined Newtonian limit.
This result is significant since it is well known that the
classical equations of General Relativity do not have a Newtonian limit either
in 2$+$1  or in 1$+$1 dimensions.
 We would like to point out that the existence of a Newtonian limit to the
 {\it semi-classical}
Einstein equation is not a unique feature of 2+1 dimension but extends to other
space-time dimensions as well. For instance, it is well known that in 1+1
dimensions
 the Einstein action is a topological invariant and consequently has no
dynamical
content. However, the semi-classical Einstein equations now give$^{27}$
$0 = 8\pi G_1 (T + \langle T \rangle )$. For conformally invariant fields
$\langle T \rangle$ is given by the trace anomaly, which in 1+1 dimensions is
simply proportional to the Ricci Scalar [Birrell and Davies$^{14}$ p.178], as a
result
the Semi-classical Einstein equations yield

$$R = 8\pi G_1 T\eqno(7.1)$$

\n which has both dynamical content as well as a Newtonian limit$^{28}$.

\n We would finally like to mention that although our analysis in this paper
has
 regarded  point sources to be fixed, it might be of interest to extend the
present
 approach
to moving sources as well. This case clearly bears a close resemblance to
moving
 mirrors --- in both cases boundary condition, instead of remaining fixed are
functions of both space and time. Consequently, as in the case of moving
mirrors
one might expect particle creation effects to be present, modifying the motion
of
 the point mass and leading to an increase in the entropy of the 2+1
dimensional
universe.

\n It would also be of interest to extend the treatment given in this paper to
2+1 dimensional space-times containing several point masses. Such a space-time
is well described by the static multi-centre metric$^{29}$~:

$$ds^2 = dt^2 - \prod_i \vert r - r_i \vert^{-8Gm_i} (dr^2 + r^2
d\t^2),\eqno(7.2)$$

\n in which particles of mass $m_i$ are located at $r_i$. (For a single point
mass,
 (7.2)can  easily be brought to the form (1.5) by a change
 of coordinates). For two masses
the problem reduces to one of finding the casimir force between two cones.
(In 3+1 dimensions, the corresponding problem would be one of determining the
casimir force between two cosmic strings.) This problem bears a close affinity
to that of the
casimir force between two wedges$^{30}$, and is presently being studied.

\bigskip


\section{Acknowledgements}

\n We benefitted from conversations with Sanjeev Dhurandhar and Jayant
Narlikar.
 One of us
(V.S.)  wishes to acknowledge stimulating discussions with Salman Habib.
T.S. wishes to acknowledge financial support from the Council of Scientific and
Industrial Research , India.



\section {Note added in Proof}


It is interesting to note, that the space-time associated with a plane domain

wall has the form [7]:
$$
ds^2 = (1 - \kappa z)^2 dt^2 - dz^2 - (1 - \kappa z)^2 \exp(2\kappa t)(dx^2 +
dy^2),
$$
where $\kappa = 2\pi G \sigma$ is the surface tension of the wall.

In the plane of the wall $(z = 0)$, this metric reduces to that of a $(2+1)$-
dimensional de Sitter space. It has been recently shown that a vacuum bubble
after nucleation also has the internal geometry of a $(2+1)$-
dimensional de Sitter space, and that perturbations on it can be described by
a scalar field with a tachyonic mass [32]. Consequently the analysis of
Sec.{\bf V},
is relevant to the study of both classical and quantum fluctuations on domain
walls and vacuum bubbles. (A conical $(2+1)$-
dimensional de Sitter space, of the kind considered in Sec.{\bf V}, would
describe
the metric on a domain wall pierced by a cosmic string.)
 From the form of the domain wall metric described above and the phenomenon of
the inversion of statistics in odd dimensional space-times discussed in
Sec.{\bf V}
and [24], it follows that a
comoving particle detector registering scalar particles and
confined to move in the $z = 0$ plane, will register a thermal Fermi - Dirac
response, characterised by a temperature
$T = {\kappa \over 2\pi}$.  On the other hand, since the $(z,t)$ part of the
above metric describes a $(1+1)$ - dimensional Rindler space, a comoving
particle detector outside of the wall, will for the same scalar particles,
register a Bose - Einstein distribution at an identical temperature. Thus
depending upon its trajectory, a comoving particle detector in the space-time
of a
domain wall, can register either a Fermi - Dirac, or a Bose - Einstein
distribution of particles.



\section{Appendix A}

\n {\bf The method of images}
\medskip

\n In $\S$2 we made the assertion

$$G_p (r, \t ; r', \t') = \sum\limits^{p-1}_{k=0} G_1 \bigg(r, \t ; r', \t' +
{2\pi k \over p}
\bigg)\eqno(A.1)$$

\n (see 2.10)

\n where

$$G_p (x, x') = {1\over 4 \pi} \sum\limits^\infty_{\scriptstyle m = -\infty
\atop
 (m \not= 0)} {X^{p \vert m \vert}\over
\vert m \vert} e^{ipm(\t - \t')}  - {p \over 2\pi} {\rm ln}~r'. \eqno(A.2)$$

\n and $X = {r \over r'} < 1.$

\n In order to prove (A.1) it is sufficient to establish that

$$\sum\limits^\infty_{\scriptstyle m = -\infty \atop (m \not= 0)} {X^{p \vert m
\vert}\over
\vert m \vert} e^{ipm(\t - \t')}  = - \sum\limits^{p-1}_{k=0}{\rm ln}~\bigg[
1 + X^2 - 2X ~{\rm cos}~ \bigg(\t - \t' + {2\pi k \over p}\bigg)
\bigg].\eqno(A.3)$$

\n To do this we rewrite the RHS of (A.3) using the expansion$^{12}$

$$-{\rm ln}~ \bigg[ 1 + X^2 - 2X ~{\rm cos}~ ( \D \t + {2\pi k \over p}) \bigg]
= \sum\limits^\infty_{\scriptstyle m = -\infty \atop m \not= 0} {X^{ \vert m
\vert}
\over\vert m \vert} e^{im( \D \t + {2\pi k \over p})} . \eqno(A.4)$$

\n Then

$$- \sum\limits^{p-1}_{k=0}{\rm ln}~ \bigg[ 1 + X^2 - 2X ~{\rm cos}~
( \D \t + {2\pi k \over p}) \bigg]
= \sum\limits^\infty_{\scriptstyle m = -\infty \atop m \not= 0} {X^{ \vert m
\vert}\over
\vert m \vert} \sum\limits^{p-1}_{k=0} e^{im( \D \t + {2\pi k \over
p})}.\eqno(A.5)$$

\n Since

$$\sum\limits^{p-1}_{k=0} e^{im {2\pi k \over p}} = p \sum\limits^\infty_{n =
-\infty}
 \d_{m,np} \eqno(A.6)$$

\n substituting (A.6) in (A.5) we get ($m \not= 0$ is assumed in A.7--A.8)

$$\sum\limits^\infty_{m = -\infty} {X^{ \vert m \vert}\over
\vert m \vert} \sum\limits^{p-1}_{k=0} e^{im( \D \t + {2\pi k \over p})}
= p \sum\limits^\infty_{m = -\infty} {X^{ \vert m \vert}\over
\vert m \vert} e^{im\D \t} \sum\limits^\infty_{n = -\infty} \d_{m,np}
\eqno(A.7)
$$

$$~~~~~~~~~~~~~~~~~~~~~~~~~~~~~ = \sum\limits^\infty_{n = -\infty}
{X^{p \vert n \vert}\over
\vert n \vert} e^{ipn\D \t}.\eqno(A.8)$$

\n which is the LHS of (A.3).

\n Having established (A.3) we note that $G_p (x, x')$ as defined in(A.2) can
be
written as

$$\eqalignno{G_p (r, \t ; r', \t') &= -{1 \over 4\pi} \sum\limits^{p-1}_{k=0}
 {\rm ln}~\bigg[
1 + X^2 - 2X ~{\rm cos}~\bigg( \D \t + {2\pi k \over p}\bigg) \bigg] - {p \over
2\pi} ln ~r'\cr
&= -{1 \over 2\pi} \sum\limits^{p-1}_{k=0}  {\rm ln}~\bigg[r^2 + r'^2 - 2rr'
{\rm cos}~\bigg( \D \t + {2\pi k \over p}\bigg) \bigg]^{1/2}\cr
&= \sum\limits^{p-1}_{k=0} G_1 \bigg( r, \t; r', \t' +
{2\pi k \over p}\bigg)&(A.9)\cr}$$

\n which is what we set out to establish. (Equivalently $G_p^{\rm reg} (x, x')
= G_p - G_1 = \sum\limits^{p-1}_{k=1} G_1 (r, \t , r' , \t' + {2\pi k \over
p})~~$)

\n Similarly the two point functions of $\S$3 and $\S$6 can also be obtained
according to the method of images~:

$$D_p ( x,  x') =  \sum\limits^{p-1}_{k=0} D_{\rm Mink} ( x ,
x_k')\eqno(A.10)$$

\n $(p = 1,2,\ldots)$ for a field satisfying periodic boundary conditions in
$\t$; and

$$D_p^T ( x, x') =  \sum\limits^{p-1}_{k=0} (-1)^k D_{\rm Mink} (x ,
x_k')\eqno(A.11)$$

\n $(p = 2,4,6,\ldots)$ for a twisted field satisfying antiperiodic boundary
condition in $\t$.

\n Here $x \equiv (t, r, \t), ~~x_k' \equiv (t', r', \t' +{2\pi k \over p} )$
and
for flat space

$$D_{\rm Mink} (x, x_k' ) = {1\over 4\pi \sigma_k}\eqno(A.12)$$

\n where

$$\sigma_k = \vert x -  x_k' \vert = \bigg[ r^2 + r'^2 - 2rr'{\rm cos}~( \D \t
+ {2\pi k \over p}) - \D t^2\bigg]^{1/2}$$

\n  To prove (A.10) and (A.11) we note that from (3.10) and (6.3)

$$D_p (x, x') = {p\over 4\pi}  {\int\limits^{\infty}_0} d\o  e^{-i\o (t - t')}
\sum\limits^\infty_{m = -\infty}
e^{ipm(\t - \t')} J_{p \vert m \vert} (\o r) J_{p \vert m \vert} (\o
r')\eqno(A.13)$$

\n and

$$D^T_p (x, x') = {p\over 4\pi}\int\limits^\infty_0   d\o  e^{-i\o (t - t')}
\sum\limits^\infty_{m = -\infty}
e^{ip (m + {1\over 2} ) (\t - \t')} J_{p \vert m  +{1\over 2}\vert} (\o r)
J_{p \vert m +{1\over 2}\vert } (\o r')\eqno(A.14)$$

\n (A.10) and (A.11) are established immediately if we use the generalised laws
of
 addition for Bessel functions (Davies and Sahni$^{9}$).

$$p \sum\limits^\infty_{m = -\infty} J_{p \vert m \vert} (\o r) J_{p \vert m
\vert}
 (\o r') e^{ipm \D \t} = \sum\limits^{p-1}_{k=0} J_0 (\o r_k ) \eqno(A.15)$$

$$r_k = \bigg[ r^2 + r'^2 -2rr' ~{\rm cos} ~(\t - \t' + {2\pi k \over
p})~\bigg]^{1/2}$$

\n $(p = 1,2,\ldots )$ for normal modes, and

$$p \sum\limits^\infty_{m = -\infty} J_{p \vert m +{1\over 2}\vert} (\o r)
 J_{p \vert m +{1\over 2}\vert} (\o r')  e^{ip (m + {1\over 2})\D \t} =
 \sum\limits^{p-1}_{k=0} (-1)^k J_0 (\o r_k)\eqno(A.16)$$

\n $(p = 2,4,6,\ldots )$ for twisted modes.

\n To establish (A.15) we use the standard summation formula for Bessel
functions$^{12}$.

$$ J_o (\o r_0 ) =  \sum\limits^\infty_{l = -\infty}  J_{ \vert l \vert} (\o r)
J_{ \vert l \vert} (\o r')
e^{il \D \t} \eqno(A.17)$$

\n Substituting (A.17) into the right hand side of (A.15) we obtain

$$\sum\limits^{p-1}_{k=0} J_o (\o r_k) =  \sum\limits^\infty_{l = -\infty}
 J_{\vert l \vert} (\o r)
J_{\vert l \vert} (\o r')
e^{il \D \t}  \sum\limits^{p-1}_{k=0} e^{il {2\pi k \over p}},\eqno(A.18)$$

\n again since

$$\sum\limits^{p-1}_{k=0} e^{il {2\pi k \over p}} = p
\sum\limits^\infty_{l = -\infty}
\d_{l,mp}\eqno(A.19)$$

\n (A.18) reduces to

$$\sum\limits^{p-1}_{k=0} J_o (\o r_k) = p  \sum\limits^\infty_{l = -\infty}
J_{p \vert m \vert} (\o r)
J_{p \vert m \vert} (\o r') e^{ipm \D \t}\eqno(A.20)$$

\n which is what we set out to establish.

\n Similarly substituting (A.17) into the right hand side of (A.16) we get

$$\sum\limits^{p-1}_{k=0} (-1)^k J_0 (\o r_k) = \sum\limits^\infty_{l =
-\infty}
J_{\vert l \vert} (\o r)
J_{\vert l \vert} (\o r') e^{il \D \t}
\sum\limits^{p-1}_{k=0}  (-1)^k e^{il {2\pi k \over p}}.\eqno(A.21)$$

\n Again noting that for $p = 2,4,6,\ldots$

$$\sum\limits^{p-1}_{k=0} e^{i\pi k ( 1 + {2l \over p})} = p \sum\limits^
\infty_{m = -\infty} \d_{l,( m + {1\over 2}) p}\eqno(A.22)$$

\n and substituting (A.22) in (A.21) we obtain (A.16)

$$\sum\limits^{p-1}_{k=0} (-1)^k J_0 (\o r_k) = p \sum\limits^\infty_{m =
-\infty}
J_{p \vert m + {1\over 2}\vert} (\o r) J_{p \vert m + {1\over 2}\vert} (\o r')
e^{ip ( m + {1\over 2})\D \t}\eqno(A.23)$$

\n Finally we would like to point out that the method of images retains its
validity for non-static space-times possessing a symmetry axis, such as the
metric

$$ds^2 = a^2 (\eta ) (d\eta^2 - dr^2 - {r^2 \over p^2} d\t^2)\eqno(A.24)$$

\n The Green's function in (A.24) has the following form (for normal modes)

$$D_p ( x, x' ) = {p\over 4\pi} {1\over \sqrt{a(\eta )
a(\eta')}}\int\limits_0^\infty
d\o ~\chi_\o (\eta ) \chi^{*}_\o (\eta ' ) \sum\limits^\infty_{m = -\infty}
 e^{ipm(\t - \t')} J_{p\vert m \vert} (\o r)
J_{p\vert m \vert} (\o r')\eqno(A.25)$$

\n where $\chi_\o (\eta )$ satisfy the time component of the Klein-Gordon
 equation$^{14}$

$${d^2 \chi_\o \over d\eta^2} + \{ \o^2 + a^2(\eta ) [ m^2 + ( \xi - {1
\over 8} ) R(\eta ) ]\} \chi_\o = 0\eqno(A.26)$$

\n and the normalisation condition $\chi_\o  \partial_\eta \chi^{*}_\o
 -\chi^{*}_\o \partial_\eta \chi_\o= i$.

\n $R(\eta)$ is the scalar curvature for the space-time (A.23), and $\eta$
is the conformal time $\eta = \int {dt \over a}$.

\n Similarly for twisted modes $(p = 2,4,6,\ldots )$

$$\eqalign{D_p^T ( x, x' ) = {p\over 4\pi} {1\over \sqrt{a(\eta )
a(\eta')}} &\int\limits_0^\infty
d\o ~\chi_\o (\eta ) \chi^{*}_\o (\eta ' )
J_{p \vert m + {1\over 2}\vert} (\o r)
J_{p \vert m + {1\over 2}\vert} (\o r') \cr & ~~~~~~~~~~
\times \sum\limits^\infty_{m = -\infty}e^{ip ( m + {1\over 2})\D \t}\cr}
 \eqno(A.27)$$

\n Clearly the proof of the image formulae established for the flat space
 Green's function
(A.13, A.14) can be extended to this case also, since the essential element in
the proof --- the generalised summation formulae for Bessel functions
 (A.15, A.16) ---
can be, with equal validity, applied to (A.25) and (A.27) resulting in

$$\eqalign{D_p ( x, x' ) &= {p\over 4\pi} {1\over \sqrt{a(\eta ) a(\eta')}}
\sum\limits^{p-1}_{k=0} \int d\o \chi_\o (\eta) \chi^{*}_\o (\eta') J_0 (\o r_k
) \cr
& = \sum\limits^{p-1}_{k=0} D_1 (x, x'_k).\cr}\eqno(A.28a)$$

\n and

$$\eqalign{D_p^T
 ( x, x' ) &= {p\over 4\pi} {1\over \sqrt{a(\eta ) a(\eta')}}
\sum\limits^{p-1}_{k=0} (-1)^k \int d\o \chi_\o (\eta) \chi^{*}_\o (\eta')
 J_0 (\o r_k ) \cr
& = \sum\limits^{p-1}_{k=0}  (-1)^k D_1 (x, x'_k).\cr}\eqno(A.28b)$$

\bigskip



\section{Appendix B}

\n {\bf The two point function  in $n$-dimensional de Sitter space}

\medskip
\n The n-dimensional spatially flat Robertson Walker line element has the form

$$ds^2 = dt^2 - a^2(t)~ dl^2 ~=~ a^2 (\eta ) [d\eta^2 - dl^2 ]\eqno(B.1)$$

\n where $dl^2 = \sum\limits^{n-1}_{i=1} dx^2_i$ and $\eta$ is the conformal
time
coordinate $\eta = \int {dt \over a(t)}.$ In the case of de Sitter space
 $a(\eta) = - {1 \over H
\eta},~-\infty \leq \eta < 0$, where $H$ is the Hubble parameter,
 $H = \sqrt {{\L \over
3}}$ and $\L$ is the n-dimensional cosmological constant.
A massive, real scalar field in (B.1)  satisfies the n-dimensional
 Klein-Gordon equation

$$(~\dal ~ + m^2 + \xi R(x) ) ~\phi (x) = 0\eqno(B.2)$$

\n where $\dal ~ = {1\over \sqrt{-g}} \partial_\m (g^\mn \sqrt{-g} \p_\nu).$

\n For purposes of quantisation, the scalar field $\phi$ may be treated as an
 operator and decomposed into modes so that

$$\hat \phi (x) = \int d^{n-1} k~ [ \hat a_k u_k (x) + \hat a^\dagger_k
 u^{*}_k (x) ]\eqno(B.3)$$

\n where $\hat a_k$ and $\hat a^\dagger_k$ are the $n-1$ dimensional
annihilation
 and creation operators. $u_k(x)$ can be written as$^{14}$

$$u_k (x) = (2\pi)^{{1-n\over 2}} a(\eta)^{{2-n\over 2}} e^{ik_i x^i}
\chi_k (\eta)\eqno(B.4)$$

\n where $k = [ ~\sum\limits^{n-1}_{i=1} k^2_i ~]^{1/2},$ and $\chi_k (\eta)$
 satisfies

$${d^2 \over d\eta^2} \chi_k (\eta) + \{ k^2 + a^2 (\eta)
[ m^2 + (\xi - \xi (n))~ R(\eta )~]
\} ~\chi_k (\eta) = 0\eqno(B.5)$$

\n $\xi (\eta) = {(n-2 ) \over 4(n-1)}$ is the conformal coupling factor
 in $n$ dimensions,
and $R(\eta )$ is the scalar curvature of the space-time.

\n In de Sitter space $R = n (n -1) H^2 = constant$ and (B.5) may be solved
exactly
to obtain

$$\chi_k (\eta) ={1\over 2} (\pi\eta)^{1/2} H_\nu^{(2)} (k \eta ), \eqno(B.6)$$

\n $\chi_k (\eta )$ are positive frequency solutions of (A.5) normalised
according
to the Wronskian condition

$$\chi_k (\eta) {\p \over \p \eta} \chi^{*}_k (\eta ) -  \chi^{*}_k (\eta )
{\p \over \p \eta} \chi_k (\eta ) = i.\eqno(B.7)$$

\n and $\nu = \bigg[ {(n-1)^2 \over 4} - {m^2 \over H^2} - n (n-1) \xi
\bigg]^{1/2}.$

\n The complete mode functions are now

$$u_k (x) = 2^{-{3\over 2}} (2\pi)^{{2-n \over 2}} (-H\eta)^{{n-2 \over 2}}
e^{ik_i x^i} H_\nu^{(2)} (k \eta ) \eqno(B.8)$$

\n The two point function $G(x, x')$ can be obtained using a mode sum
approach$^{14}$

$$G(x, x') = \int d^{n -1} k ~u_k (x) u^{*}_k (x')\eqno(B.9)$$

\n which in this case leads to

$$G(x, x') = {1\over 8} \bigg(-{H \over 2\pi}\bigg)^{n-2} (\eta
\eta')^{{n-1\over 2}}
\int d^{n-1}k H_\nu^{(1)} (k\eta ) H^{(2)}_\nu (k\eta') e^{ik_i \D x^i}
\eqno(B.10)$$

\n where $\D x^i = x^i - x'^i$.

\n The $d^{n-1} k$ integration is carried out in polar coordinates using the
following expression for the integration over the solid angle$^{12}$

$$\int d \O_{m-1} e^{i \vec k . \D \vec x} = {(2\pi)^{m/2} J_{(m-2)/2} (k \D
x)\over
(k \D x)^{m-2 \over 2}}\eqno(B.11)$$

\n where $k = \vert \vec k \vert$ and $\D x = \vert \D \vec x \vert$. Rewriting
the Hankel functions $H_\nu (k\eta)$ in terms of the McDonald functions
 $K_\nu (k \eta)$

$$\eqalign{H_\nu^{(2)} (k\eta) &= {2\over \pi} K_\nu (-i k\eta)\cr
H_\nu^{(2)*} (k\eta') &= {2\over \pi} K_\nu (i k\eta')\cr}\eqno(B.12)$$

\n we obtain $G (x, x')$ as an integral over $k (= \vert \vec k \vert)$.

$$G (x, x') = {(-H)^{n-2}\over 2^{(n-1)/2}} {1 \over \pi^{(n+1)/2}}
{(\eta \eta')^{{n-1\over 2}}\over (\D x)^{n-3 \over 2}} \int\limits^\infty_0
dk k^{{n-1\over 2}} K_\nu (-ik\eta ) K_\nu (ik\eta') J_{{n-3\over 2}} (k \D x)
\eqno(B.13)$$

\n The above integral can be evaluated in terms of Legendre functions$^{12}$
so that, finally

$$G (x, x') = {(-H)^{n-2}\over 2 (2\pi)^{n/2}} {\G \bigg({n-1\over 2} + \nu
\bigg)
 \G  \bigg({n-1\over 2} - \nu \bigg) \over (u^2 - 1)^{{n-2 \over 4}}}
P_{\nu - {1\over 2}}^{- {(n-2)\over 2}} (u)\eqno(B.14)$$

\n where $u = {\D x^2 - \eta^2 -\eta'^2 \over 2\eta \eta'}$.

\n  $G(x, x')$ in equation (B.14) can be rewritten in terms of a hypergeometric
function$^{20}$ as

$$G (x, x') = {(-H)^{n-2}\over (4 \pi)^{n/2}} {\G \bigg({n-1\over 2} + \nu
\bigg)
\G  \bigg({n-1\over 2} - \nu \bigg) \over \G (n/2)} F\bigg(
{n-1 \over 2} + \nu, {n-1 \over 2} - \nu ; {n \over 2} ; \o \bigg)\eqno(B.15)$$

\n where $\o = 1 - {\D x^2 - \D\eta^2 \over 4 \eta \eta'}$.

\n One can easily verify that the propagator $D (x, x')$, for a massless and
 conformally coupled  scalar field, $( m = 0,~ \xi = \xi (n))$ scales
conformally
with the n-dimensional Minkowski space propagator $D_{\rm Mink} (x, x')$.
Substituting $\nu = {1\over 2}$ (corresponding to $m = 0$ and $\xi = \xi (n) )$
 in (B.15) and using
the relation$^{20}$

$$F(a, b; b; z) = {1\over (1-z)^a }\eqno(B.16)$$

\n we obtain

$$\eqalign{D (x, x') &= {(-H\eta )^{{n-2\over 2}}  (-H\eta')^{{n-2\over 2}}
\over (4 \pi)^{n/2}} {\G \bigg({n-2 \over 2}\bigg)\over [ \D x^2 - \D \eta^2 ]
^{(n-2)/2}}.\cr
&=[a (\eta) a (\eta') ] ^{{2-n\over 2}} D_{\rm Mink} (x, x'),\cr}
\eqno(B.17)$$

\n as expected.

\n In four dimensions (B.15) assumes the well known form$^{21}$

$$G (x, x') = {H^2\over 16\pi}\bigg( {1\over 4 }- \nu^2 \bigg) {\rm sec}~
\pi \nu ~F \bigg(
{3\over 2} + \nu, {3\over 2} - \nu; 2; \o \bigg). \eqno(B.18)$$

\n In three dimensions (B.15) reads

$$G (x , x') = {-H \over 4\pi} \nu {\rm cosec}~\pi\nu
{}~F\bigg(1+ \nu ,1- \nu;{3 \over 2} ; \o \bigg),\eqno(B.19)$$

\n which is used in $\S$5 to construct the scalar field propagator
in the conical de Sitter metric.

\bigskip


\section{Appendix C}

\n We outline a few steps involved in the evaluation of the integrals $s_1
(p)$,
 $s (p)$ and $s^T_1 (p)$ (3.18, 6.7b, 3.19),  for integer $p$ to get the
summations
(3.22, 6.12).

\n The substitution $z = e^x$ brings these integrals to a form suitable for
contour
integration over the contour ${\cal C}$ shown in Fig. 5.

$$I_1 = {2\over \pi} \oint\limits_{\cal C} {dz \over (z^2 -1) } \bigg[ p
{(z^{2p} + 1 )\over (z^{2p} -1)}
-{(z^2+1)\over (z^2 -1)}\bigg]\eqno(C.1a)$$

$$I_2 = {8\over \pi} \oint\limits_{\cal C} {dz \over (z^2 -1) }\bigg[ {z^2 + 1
\over (z^2 -1)^3}
-p^3 {(z^{2p} + 1) \over (z^{2p} - 1)^3} \bigg]\eqno(C.1b)$$

$$I_3 = {4\over \pi} \oint\limits_{\cal C} {dz \over (z^2 -1) } {z^p \over
(z^{2p} -1)}
{}~~~~(p~{\rm even})\eqno(C.1c)$$

\n The poles of all the contour integrals (C.1) are the $(2p)^{\rm th}$ roots
 of unity, of which the poles $e^{{ik\pi \over p}}, k = 1,\ldots, p -1$
lie within the contour
${\cal C}$ (Figure 5).

\n Evaluating (C.1) using the residue theorem we obtain (omitting lengthy
intermediate steps)

$$s_1(p) = \sum\limits^{p-1}_{k=1} ~{\rm cosec}~ {k\pi \over p}\eqno(C.2a)$$

$$s(p) = \sum\limits^{p-1}_{k=1} (~{\rm cosec}^3~ {k\pi \over p}
- {1\over 2} ~{\rm cosec}~ {k\pi \over p})\eqno(C.2b)$$

$$s^T_1(p) = \sum\limits^{p-1}_{k=1}  (-1)^k ~{\rm cosec}~ {k\pi \over p}
 + {2\over \pi}.\eqno(C.2c)$$


\vfill\eject


\section{References}

\item{$^1$}D. Kramar, H. Stephani, E. Herlt and M. MacCullum ``Exact Solutions
of
Einsteins Field Equations'', (Cambridge University Press, Cambridge 1980).

\s

\item{$^2$}S. Deser, in ``Fields, Strings and Gravity'', edited by H. Guo, Z.
Qiu
and H. Tye; CCAST (World Laboratory) Symposium/Workshop proceedings vol.6
(Gordon and Breach, New York, 1990).

\s

\item{$^3$}S. Deser, R. Jackiw and S. Templeton, Ann. Phys. (N.Y.) {\bf 140},
372 (1982).

\s

\item{$^4$}M.E. Ortiz, Nucl. Phys. {\bf 363}, 185 (1991).

\s

\item{$^5$}S. Habib, ``Some Aspects of three dimensional gravity'', University
of
Maryland (1984) (unpublished).

\s

\item{$^6$}In Einstein gravity this form of the metric is valid everywhere
outside
the point source. It is also interesting to note that the exterior metric of an
extended static source also has the asymptotically conical form (1.5) as
recently demonstrated in N.J. Cornish and N.E. Frankel, Phys. Rev. D {\bf 43},
2555 (1991).

\s

\item{$^7$}A. Vilenkin, Phys. Reports {\bf 121}, 263 (1985).

\s

\item{$^8$}G. 't Hooft, Commun. Math. Phys. {\bf 117}, 685 (1988).

\s

\item{}S. Deser and R. Jackiw, Commun. Math. Phys. {\bf 118}, 495 (1988).
\s

\item{}A. Cappelli, M. Ciafaloni and P. Valtancoli, preprint CERN-TH-6093/91.

\s

\item{$^9$}Quantum effects near cosmic strings were investigated in~:

\s

\item{}T.M. Helliwell and D.A. Konkowski, Phys. Rev. D {\bf 13}, 1918 (1986).

\s

\item{}B. Linet, Phys. Rev. D {\bf 33}, 1833 (1986).

\s

\item{}B. Linet, Phys. Rev. D {\bf 35}, 536 (1987).

\s

\item{}A.G. Smith, Tufts Univeristy preprint (1986).

\s

\item{}A.G. Smith, In ``The Formation and Evolution of Cosmic Strings', Edited
by G.W. Gibbons, S.W. Hawking and T. Vachaspati (Cambridge University Press,
Cambridge 1990).

\s

\item{}J.S. Dowker, J. Phys. {\bf A10}, 115 (1977).

\s

\item{}J.S. Dowker, Class. Quant. Grav. {\bf 4}, L157 (1987).

\s

\item{}J.S. Dowker, In ``The Formation and Evolution of Cosmic Strings'',
Edited
by G.W. Gibbons, S.W. Hawking and T. Vachaspati (Cambridge University Press,
Cambridge 1990).

\s

\item{}P.C.W. Davies and V. Sahni, Class. Quant. Grav. {\bf 5}, 1 (1987).

\s

\item{}V.P. Frolov and E.M. Serebriany, Phys. Rev. D {\bf 34}, 2840 (1986).

\s

\item{}L. Parker, Phys. Rev. Letts. {\bf 59}, 1369 (1987).

\s

\item{}V. Sahni, Mod. Phys. Lett. A {\bf 3}, 1425 (1988).

\s

\item{}W.A. Hiscock, Phys. Lett B {\bf 188}, 317 (1987).

\s

\item{}I.H. Russell and D.J. Toms, Class. Quant. Grav. {\bf 6}, 1343 (1989).

\s

\item{}A. Sarmiento and S. Hacyan, Phys. Rev. D {\bf 38}, 1331 (1988).

\s

\item{$^{10}$}We refer the reader to the paper by B. Linet and A.G. Smith (ref.
9)
where a similar analysis has been carried out for cosmic strings.

\s

\item{$^{11}$}The method of images has also been discussed in the context of
cosmic
strings by A.G. Smith and by P.C.W. Davies and V. Sahni (ref. 9).

\s

\item{$^{12}$}I.S. Gradshteyn and I.M. Ryzhik, ``Tables of Integrals, Series
and
 Products'' (New York~: Academic Press 1980).

\s

\item{$^{13}$}B.S. DeWitt, Phys. Reports {\bf 19C}, 295 (1975).

\s

\item{}S.M. Christensen, Phys. Rev. D {\bf 14}, 2490 (1976).

\s

\item{$^{14}$}N.D. Birrell and P.C.W. Davies, ``Quantum Fields in Curved
Space''
(Cambridge University Press, Cambridge 1982).

\s

\item{$^{15}$}T.M. Helliwell and D.A. Konkowski (ref. 9) also obtain similar
relationships between the various derivatives of the Green's function in the
space-time of a cosmic string.

\s

\item{$^{16}$}Dimensionally $G_2 \equiv G_3 l^{-1}$, where $G_3$ is the
gravitational
constant in 3+1 dimensions and $l$ is a fundamental length scale --- say the
Planck length.

\s

\item{$^{17}$}Our treatment in this section closely follows that of W.A.
Hiscock (ref. 9), who evaluated the semi-classical correction to the cosmic
string metric.

\s

\item{$^{18}$}L.D. Landau and E.M. Lifshitz, ``The Classical Theory of
Fields'', p. 295
,(Pergamon Press Ltd. 1975).

\s

\item{$^{19}$}J.D. Barrow, A.B. Burd and D. Lancaster, Class. Quantum. Grav.
{\bf 3}, 551 (1986). See also N.J. Cornish and N.E. Frankel (ref.6).

\s

\item{$^{20}$}M. Abramowitz and I.A. Stegun, ``Handbook of Mathematical
Functions''
\item{}(Dover, New York, 1970).

\s

\item{$^{21}$}T.S. Bunch and P.C.W. Davies, Proc. R. Soc. London {\bf A360},
117 (1978).
\item{} An alternate derivation of (B.16) is given in P. Candelas and
D.J.Raine,
Phys. Rev. D{\bf 12}, 965 (1975) and
J.S. Dowker and R. Critchley , Phys. Rev. D {\bf 13 }, 3224 (1976).

\s

\item{$^{22}$}B. Allen and A. Folacci, Phys. Rev. D {\bf 35}, 3771 (1987).

\s

\item{$^{23}$}W.G. Unruh, Phys. Rev. D {\bf 14}, 870 (1976).

\s

\item{}B.S. DeWitt, in ``General Relativity'', Editors S.W. Hawking and W.
Israel
(Cambridge University Press, Cambridge 1979).

\s

\item{$^{24}$}S. Takagi, Prog. Theor. Phys. Suppl. No. 88 (1986).

\s

\item{$^{25}$}C.J. Isham, Proc. Royal Soc. London {\bf A362}, 383 (1978).

\s

\item{}C.J. Isham, Proc. Royal Soc. London {\bf A364}, 591 (1978).

\s

\item{$^{26}$}L.H. Ford, Phys. Rev. D {\bf 21}, 949 (1980).

\s

\item{$^{27}$}T.S. Bunch and P.C.W. Davies, J. Phys. {\bf A 11}, 1315 (1978)

\s

\item{}N. Sanchez, Nucl. Phys. {\bf B266}, 487 (1986).

\s

\item{$^{28}$}$G_1$ should be veiwed as an `induced' gravitational
constant ( \`a la Sakharov) since its value is determined solely by one-loop
quantum gravitational effects.
Equation (7.1) can also be derived from a local action principle as
demonstrated by~: C. Teitleboim in ``Quantum Theory of Gravity'', ed. S.
Christensen
(Adam Hilger, Bristol 1984) p.327; and R. Jackiw, {\it ibid}, p.403.

\item{} A discussion
on the Newtonian limit of (7.1) can also be found in N.J. Cornish and N.E.
Frankel
(ref. 6), and in ~: J. Gegenberg, P.F. Kelly, R.B. Mann and D. Vincent, Phys.
Rev. D
{\bf 37}, 3463 (1988)

\s

\item{$^{29}$}S. Deser, R. Jackiw and G. 't Hooft, Ann. Phys. (N.Y.) {\bf 152},
220
(1984).

\s

\item{$^{30}$} D. Deutch and P. Candelas, Phys. Rev. D {\bf 20} , 3063, (1979).

\vfill\eject


\section{Figure Captions}

\item{Fig. 1}The equipotential contours of the electrostatic force
field of a point charge $Q$ are shown for a space-time possessing a deficit
angle ${3 \pi \over 2}$ using the method of images (see $\S$2 and Appendix A).

\medskip

\item{Fig. 2}The $p$ dependence of $8 \pi r \langle \phi^2 (r)\rangle $ is
shown
for both twisted $~(\equiv s_1^T (p)~)~$ and untwisted fields $~(~\equiv s_1
(p))$.

\medskip

\item{Fig. 3}The $p$ dependence of the vacuum energy densities $r^3 \langle
T_{00}
\rangle$ and $r^3\langle T_{00} \rangle^T$ (see (3.21) and  (6.9) ) is shown
for
 untwisted  --- solid lines,
and twisted scalar fields --- dashed lines. Two values of the coupling
parameter
 $\xi$ are
considered~:  minimal coupling
$(\xi = 0)$, and conformal coupling $(\xi = 1/8)$.

\medskip

\item{Fig. 4}The dependence of the {\it gravitating mass} $M_G$, defined in
(4.16)
 for the conformally coupled case $\xi = {1\over 8}$, is plotted against the
 mass $M$  of the point source. $M$ is related to
the deficit angle of the space-time through $\D \varphi_{\rm def} = 8\pi M$.
($M_G$ and $M$ are both expressed in units of the planck mass).

\medskip

\item{Fig. 5}The contour ${\cal C}$ and poles of (3.18, 6.7b, 3.19)
used to establish (3.22, 6.12) by means of contour integration in
Appendix C, are shown in the complex ${\cal Z}$ plane for $p = 8 $.

\bye